\documentclass[10pt]{article}
\usepackage{amssymb}
\usepackage{amsmath}
\usepackage{amsthm}
\usepackage{latexsym}
\usepackage[dvips]{epsfig}
\usepackage{mathrsfs}
\usepackage{eufrak}
\usepackage{bm}
\usepackage{authblk}

\theoremstyle{plain}
\newtheorem{proposition}{Proposition}
\newtheorem{lemma}{Lemma}
\newtheorem{theorem}{Theorem}

\newtheorem{corollary}{Corollary}

\font\SYM=msbm10
\newcommand{\Real}{\mbox{\SYM R}}

\newcommand{\Sphere}{\mbox{\SYM S}}

\font\tenscr=rsfs10 scaled1100
\font\sevenscr=rsfs7 
\font\fivescr=rsfs5 
\skewchar\tenscr='177
\skewchar\sevenscr='177
\skewchar\fivescr='177
\newfam\scrfam
\textfont\scrfam=\tenscr
\scriptfont\scrfam=\sevenscr
\scriptscriptfont\scrfam=\fivescr

\def\scri{{\fam\scrfam I}}

\newcommand{\tensor}[3]{_{#1\phantom{#2}#3}^{\phantom{#1}#2}}
\newcommand{\half}{\frac{1}{2}}

\newcommand{\er}{{\underline{r}}}
\newcommand{\es}{{\underline{s}}}

\newcommand{\D}{D}

\newcounter{mnote}

\begin{document}


\title{\textbf{The extended Conformal Einstein field equations with matter: the Einstein-Maxwell field}}

\author[,1,2]{Christian L\"ubbe \footnote{E-mail address:{\tt c.luebbe@qmul.ac.uk}, {\tt cl242@le.ac.uk}}}
\author[,1]{Juan Antonio Valiente Kroon \footnote{E-mail address:{\tt j.a.valiente-kroon@qmul.ac.uk}}}
\affil[1]{School of Mathematical Sciences, Queen Mary, University of London,
Mile End Road, London E1 4NS, United Kingdom.}
\affil[2]{Department of Mathematics, University of Leicester, University Road, LE1 8RH, United Kingdom.}

\maketitle

\begin{abstract}
A discussion is given of the conformal Einstein field equations coupled with matter whose
energy-momentum tensor is trace-free. These resulting equations are
expressed in terms of a generic Weyl connection. The article shows how
in the presence of matter it is possible to construct a conformal
gauge which allows to know \emph{a priori} the location of the
conformal boundary. In vacuum this gauge reduces to the so-called
conformal Gaussian gauge. These ideas are applied to obtain: (i) a new
proof of the stability of Einstein-Maxwell de Sitter-like spacetimes;
(ii) a proof of the semi-global stability of purely radiative
Einstein-Maxwell spacetimes. 
\end{abstract}


\section{Introduction}
\label{Introduction}

The Einstein conformal field equations are a powerful tool to prove
statements concerning the stability of vacuum spacetimes ---see
e.g. \cite{Fri86b}. These methods have been extended to deal with the
case of the gravitational field coupled to the Maxwell and Yang-Mills
fields \cite{Fri91}. In \cite{Fri95} a more general version of the
vacuum conformal field equations has been developed. These extended
conformal field equations are written in terms of a Weyl
connection. The extra gauge freedom incorporated in this
representation of the equations allows the construction of gauge
systems based on conformal structures of the spacetime. As it so often
happens in this type of considerations, a judicious gauge choice based
on geometrical considerations can greatly simplify the analysis in
question. An example of the gauge choices that can be employed are the
conformal Gaussian gauge systems introduced in \cite{Fri95}
---see as well \cite{Fri03a,Fri03c}. The extended conformal field
equations in conjunction with conformal Gaussian systems have been
used, among other things: to provide an existence proof of anti-de
Sitter spacetimes \cite{Fri95}; to construct a representation of spatial
infinity allowing for a regular finite initial value problem at
spatial infinity \cite{Fri98a}; to provide a new proof of the global
stability of the de Sitter spacetime and the semi-global stability of
Minkowski spacetime \cite{LueVal09a}; and to provide a semi-global
stability result of purely radiative vacuum spacetimes
\cite{LueVal09b}.

\smallskip
The common feature in the applications described in the previous
paragraph is that one is, ultimately, concerned with solutions to the
vacuum Einstein field equations (with or without a cosmological
constant). A key feature of conformal Gaussian systems in vacuum
spacetimes is that the conformal geodesics upon which they are constructed
render a canonical conformal factor which provides \emph{a priori}
knowledge about the location of the conformal boundary of the
spacetime. This property is, however, lost if one considers conformal
geodesics on non-vacuum spacetimes. In this article we show it is
possible to get around this difficulty if one considers a more general
class of \emph{conformal curves} to construct gauge systems. As in the
case of conformal geodesics in vacuum spacetimes, the conformal curves
provide again a canonical conformal factor which is known prior to
evolution.

\smallskip
As an application of the ideas described in the previous paragraph, in
this article we will consider initial value problems for spacetimes
$(\tilde{\mathcal{M}},\tilde{g}_{\mu\nu})$ with cosmological constant
$\lambda $ satisfying the Einstein-Maxwell field equations
\begin{subequations}
\begin{eqnarray}
&& \tilde{R}_{\mu\nu}-\frac{1}{2}\tilde{g}_{\mu\nu}\tilde{R}= -\lambda \tilde{g}_{\mu\nu} + \tilde{T}_{\mu\nu}, \label{EM-FE:1}\\
&& \tilde{T}_{\mu\nu} =\tilde{F}_{\mu\lambda} \tilde{F}^\lambda{}_\nu -\frac{1}{4}\tilde{g}_{\mu\nu} \tilde{F}_{\lambda\rho}\tilde{F}^{\lambda\rho} \label{EM-FE:2} \\
&& \tilde{\nabla}^\mu \tilde{F}_{\mu\nu}=0, \quad \tilde{\nabla}_{[\mu} \tilde{F}_{\nu\lambda]}=0, \label{EM-FE:3}
\end{eqnarray}
\end{subequations}
where $\tilde{\nabla}$ denotes the Levi-Civita connection of the
metric $\tilde{g}_{\mu\nu}$, $\tilde{R}_{\mu\nu}$, $\tilde{R}$ are the
associated Ricci tensor and Ricci scalar, and $\tilde{F}_{\mu\nu}$
denotes the Maxwell tensor ---the conventions for the geometric
quantities used above, will be set out in detail in
Section \ref{conventions}. The discussion of the solutions to equations
\eqref{EM-FE:1}-\eqref{EM-FE:2} will be carried out in terms of a
conformally rescaled, unphysical metric $g_{\mu\nu}$ related to the physical metric
$\tilde{g}_{\mu\nu}$ according to
\begin{equation}
\label{rescaled:metric}
g_{\mu\nu} = \Theta^2 \tilde{g}_{\mu\nu}.
\end{equation}

\medskip
The gauge systems based on the new class of conformal curves are used to provide a new
(simpler) proof of the existence and stability of Einstein-Maxwell de
Sitter-like spacetimes. We also provide a stability proof of purely
radiative Einstein-Maxwell spacetimes. These particular applications
lead us to consider the \emph{extended Einstein conformal field
equations with matter}. To the best of our knowledge, this is the
first time these equations are considered. As a simplifying technical
assumption, our general considerations will be restricted to matter
models with a trace-free stress-energy tensor ---a property satisfied
by the electromagnetic field. Although the particular examples to be
considered are only concerned with the Einstein-Maxwell equations,
most of our discussion can be adapted to trace-free perfect fluids
(sometimes also called conformal fluids) ---this will be discussed in
future work.

\subsection*{Outline of the article}
We start by summarising our conventions and
the basic ideas behind the notion of conformal rescaling
in Section \ref{conventions}. The conventions follow 
closely those used in references \cite{LueVal09a,LueVal09b}. 
Section \ref{Section:WeylConnections} presents a brief review 
of the notion of Weyl connection and the transformation 
formulae for the connection and the Schouten tensor. 
Section \ref{Section:XCFE} gives the formulation of the
extended conformal field equations with matter in both a frame and a
spinorial formalism. 
Section \ref{Section:ConformalCurves} introduces
the concept of conformal curves and the associated generalised
conformal Gaussian systems. These gauge systems are instrumental in our
subsequent analysis as combined with the extended conformal field
equations, they render hyperbolic reductions for which the location of
the conformal boundary is know \emph{a priori}. 
Section \ref{Section:HyperbolicReductions} provides a discussion of the
procedure of hyperbolic reduction for the geometric part of the
extended conformal field equations in generalised Gaussian
systems. 
Section \ref{Section:MaxwellEquations} is concerned with the 
matter part of the field equations, which in the case under 
consideration is given by the Maxwell field. 
Section \ref{Section:StructuralProperties} summarises the key 
structural properties of the evolution equations implied by the 
conformal field equations with a view to applications involving 
existence and stability results. 
Section \ref{Section:PropagationOfConstraints} discusses the 
so-called propagation of the constraints. 
Section \ref{Section:deSitterSpacetimes} is concerned with the 
first application of the methods developed in the article: a 
new proof of the stability of Einstein-Maxwell spacetimes which 
have a global structure similar to that of the de Sitter spacetime. 
Finally, Section \ref{Section:RadiativeSpacetimes} provides a
second application: a stability result for Einstein-Maxwell radiative
spacetimes. This result generalises the analysis for the purely vacuum
case carried out in \cite{LueVal09b}.

\section{Basics and conventions}
\label{conventions}

\subsection{The curvature of the physical spacetime manifold}

Throughout this article we work with a spacetime
$(\tilde{\mathcal{M}},\tilde{g}_{\mu\nu})$, where
$\tilde{g}_{\mu\nu}$, ($\mu,\nu=0,1,2,3$) is a Lorentzian metric with
signature $(+,-,-,-)$. We will denote by $\tilde{\nabla}$ the
Levi-Civita connection of $\tilde{g}_{\mu\nu}$ ---that is, the unique
torsion-free connection that preserves the metric
$\tilde{g}_{\mu\nu}$. As in the introduction, let
$\tilde{R}_{\mu\nu\lambda\rho}$, $\tilde{R}_{\mu\nu}$ and $\tilde{R}$
denote, respectively, the Riemann curvature tensor, the Ricci tensor
and the Ricci scalar of the Levi-Civita connection
$\tilde{\nabla}$. The conventions for the curvature used in this
article are such that
\begin{equation}
\label{Riemann:convention}
\tilde{R}^\mu{}_{\nu\lambda\rho} \xi^\nu =\left(\tilde{\nabla}_\lambda \tilde{\nabla}_\rho -\tilde{\nabla}_\rho \tilde{\nabla}_\lambda \right)\xi^\mu, \quad\quad
\tilde{R}_{\mu\nu}=\tilde{R}^\alpha{}_{\nu\alpha\mu}, \quad\quad
\tilde{R} = \tilde{R}_{\mu\nu} \tilde{g}^{\mu\nu} .
\end{equation}
For the Riemann tensor one has the decomposition
\begin{equation}
\tilde{R}^\mu{}_{\nu\lambda\rho} = \tilde{C}^\mu{}_{\nu\lambda\rho}+ 2\left( \delta^\mu{}_{[\lambda} \tilde{P}_{\rho]\nu} - \tilde{g}_{\nu[\lambda} \tilde{P}_{\rho]\sigma}\tilde{g}^{\sigma\mu}\right),
\label{DecompositionPhysicalRiemann}
\end{equation}
where $\tilde{C}^\mu{}_{\nu\lambda\rho}$ denotes the \emph{conformal
Weyl tensor} of $\tilde{g}_{\mu\nu}$, while the trace parts are given
in terms of the \emph{Schouten tensor} defined by
\[
\tilde{P}_{\mu\nu} \equiv \tfrac{1}{2} \left(\tilde{R}_{\mu\nu} - \tfrac{1}{6}\tilde{R} \tilde{g}_{\mu\nu}\right)
\]
In terms of an arbitrary stress-energy tensor $\tilde{T}_{\mu\nu}$, it is given by
\begin{equation}
\tilde{P}_{\mu\nu} =\tfrac{1}{2} \tilde{T}_{\mu\nu} - \tfrac{1}{6}\tilde{T}_{\rho\sigma}\tilde{g}^{\rho\sigma} \tilde{g}_{\mu\nu} + \lambda \tilde{g}_{\mu\nu}, \quad \lambda \equiv \tfrac{1}{6}\tilde{\lambda}.
\label{Schouten:T}
\end{equation}

\subsection{Conformal rescalings}
Let $\tilde{g}_{\mu\nu}$ and $g_{\mu\nu}$ be two Lorentzian metrics
which are conformally related according to equation
\eqref{rescaled:metric}. Let $[\tilde{g}]$ denote the conformal 
class of $\tilde{g}_{\mu\nu}$. Two invariants of the conformal class are the tensor
\begin{equation}
\label{DefinitionS}
  S_{\mu\nu}{}^{\lambda \rho} = \delta_\mu{}^\lambda \delta_\nu{}^\rho + \delta_\nu{}^\lambda \delta_\mu{}^\rho - \eta_{\mu\nu} \eta^{\lambda \rho}.
\end{equation}
and the conformal Weyl tensor
\[
\tilde{C}^\mu{}_{\nu\lambda\rho} = C^\mu{}_{\nu\lambda\rho}.
\]
The Levi-Civita covariant derivative of the metric $g_{\mu\nu}$ 
will be denoted by $\nabla$. In the sequel, 
it will be convenient to consider a frame $e_k$, $k=0,1,2,3$ which is
orthonormal with respect to the metric $g_{\mu\nu}$.  That is,
\begin{equation}
\label{frame:metric}
g(e_i,e_j) =\eta_{ij}.
\end{equation}
In what follows frame components are always taken with respect to the frame $e_k$. In particular, $\nabla_{i}$,
$\tilde{\nabla}_{i}$ will denote the covariant derivatives in the
direction of $e_i$. Let 
\[
\Upsilon_i \equiv \Theta^{-1} \nabla_i \Theta.
\]
Furthermore, let $\Gamma_i{}^j{}_k$, $\tilde{\Gamma}_i{}^j{}_k$ denote
the connection coefficients of $\nabla$, $\tilde{\nabla}$ with respect
to the frame $e_i$.  One has that
\[
\Gamma_i{}^j{}_k - \tilde{\Gamma}_i{}^j{}_k = S_{ik}{}^{jl}\Upsilon_l.
\]
An analogous decomposition to that of equation
\eqref{DecompositionPhysicalRiemann} holds for the Riemann tensor
$R_{\mu\nu\lambda\rho}$.

\section{Weyl connections}
\label{Section:WeylConnections}

In this article, we will also consider connections $\hat{\nabla}$ (not
necessarily Levi-Civita) which respect the conformal structure of the
 conformal class $[\tilde{g}]$, in the sense that
\begin{equation}
\label{nabla_hat_of_g}
\hat{\nabla}_\lambda \tilde{g}_{\mu\nu}= - 2 b_\lambda \tilde{g}_{\mu\nu}, \quad 
\hat{\nabla}_\lambda g_{\mu\nu}= - 2 f_\lambda g_{\mu\nu},
\end{equation}
 
for some 1-forms $b_\mu$ and $f_\mu$. One has that
\begin{subequations}
\begin{eqnarray}
&& \hat{\Gamma}_\mu{}^\lambda{}_\nu - \tilde{\Gamma}_\mu{}^\lambda{}_\nu = S_{\mu \nu}{}^{\lambda \rho}
b_\rho, \label{ChangeOfConnection:1}\\
&& \hat{\Gamma}_\mu{}^\lambda{}_\nu - \Gamma_\mu{}^\lambda{}_\nu = S_{\mu \nu}{}^{\lambda \rho}
f_\rho. \label{ChangeOfConnection:2}
\end{eqnarray}
\end{subequations}
We shall write the above equations as
\[
\hat{\nabla} - \tilde{\nabla} = S(b), \quad \hat{\nabla} - \nabla=S(f).
\]

The fact that $\tilde{g}_{\mu\nu}$ and $g_{\mu\nu}$ are assumed to be
conformally related implies
\[
b_\mu = \Upsilon_\mu + f_\mu.
\]

The Riemann and Ricci tensors of the Weyl connection $\hat{\nabla}$
are defined in an analogous way to (\ref{Riemann:convention}) and will be
denoted by $\hat{R}_{\mu\nu\lambda\rho}$ and $\hat{R}_{\mu\nu}$
respectively. The analogue of the decomposition
\eqref{DecompositionPhysicalRiemann} is given by
\begin{eqnarray*}
&& \hat{R}^\mu{}_{\nu\lambda\rho} = C^\mu{}_{\nu\lambda\rho}+ 2\left(
  \delta{}^\mu{}_{[\lambda} \hat{P}_{\rho]\nu} - \delta{}^\mu{}_{\nu}
  \hat{P}_{[\lambda\rho]} - g_{\nu[\lambda}
  \hat{P}_{\rho]\sigma}g^{\sigma\mu}  \right),  \\
&& \phantom{\hat{R}^\mu{}_{\nu\lambda\rho}}= C^\mu{}_{\nu\lambda\rho}+ 2 S_{\nu[\lambda}{}^{\mu\sigma}\hat{P}_{\rho]\sigma},
\end{eqnarray*}
where the Schouten tensor of $\hat{\nabla}$, denoted by
$\hat{P}_{\mu\nu}$, is given by
\[
\hat{P}_{\mu\nu} \equiv \tfrac{1}{2}\left( \hat{R}_{\mu\nu} -\tfrac{1}{2}\hat{R}_{[\mu\nu]} - \tfrac{1}{6} g_{\mu\nu} \hat{R}_{\rho\lambda} g^{\rho\lambda}  \right).
\]
Alternatively, the latter decompositions could have been written using the
physical metric $\tilde{g}_{\mu\nu}$.

\medskip
Transformation rules between the curvature tensors of the Weyl
connection $\hat{\nabla}$ and the Levi-Civita connections
$\tilde{\nabla}$, $\nabla$ can be found in \cite{Fri03a}. Important
for the subsequent discussion is the transformation rule for the
Schouten tensor. This is given by
\begin{subequations}
 \begin{eqnarray} 
 &&\tilde{P}_{\mu\nu} - \hat{P}_{\mu\nu} 
 = \tilde{\nabla}_\mu b_\nu - \tfrac{1}{2}  S
 \tensor{\mu\nu}{\rho\lambda}{}b_\rho b_\lambda \label{SchoutenPhysicalToWeyl1} \\ 
 &&\phantom{\tilde{P}_{\mu\nu} - \hat{P}_{\mu\nu}}= \hat{\nabla}_\mu b_\nu + \tfrac{1}{2}  S
 \tensor{\mu\nu}{\rho\lambda}{}b_\rho b_\lambda
 . \label{SchoutenPhysicalToWeyl2}
 \end{eqnarray}
\end{subequations}
A similar  expression holds between the tensors $P_{\mu\nu}$ and
  $\hat{P}_{\mu\nu}$ by replacing $b_\mu$ with $f_\mu$, namely:
\begin{subequations}
 \begin{eqnarray} 
 &&P_{\mu\nu} - \hat{P}_{\mu\nu} 
 = \nabla_\mu f_\nu - \tfrac{1}{2}  S
 \tensor{\mu\nu}{\rho\lambda}{}f_\rho f_\lambda \label{SchoutenUnphysicalToWeyl1} \\ 
 &&\phantom{P_{\mu\nu} - \hat{P}_{\mu\nu}}= \hat{\nabla}_\mu f_\nu + \tfrac{1}{2}  S
 \tensor{\mu\nu}{\rho\lambda}{}f_\rho f_\lambda
 . \label{SchoutenUnphysicalToWeyl2}
 \end{eqnarray}
\end{subequations}

Finally, we introduce the Cotton-York tensor associated to the connection $\hat{\nabla}$
\[
\hat{Y}_{\mu\nu\lambda} \equiv \hat{\nabla}_{\mu}\hat{P}_{\nu\lambda} -
\hat{\nabla}_{\nu}\hat{P}_{\mu\lambda}
\]
The physical Cotton-York tensor can be expressed in terms of $\tilde{T}_{\mu\nu}$ and $\tilde{T}_{\mu\nu}\tilde{g}^{\mu\nu}$
\[
\tilde{Y}_{\mu\nu\lambda} \equiv \tilde{\nabla}_{\mu}\tilde{P}_{\nu\lambda} -
\tilde{\nabla}_{\nu}\tilde{P}_{\mu\lambda} = \tilde{\nabla}_{[\mu}\tilde{T}_{\nu]\lambda}
- \tfrac{1}{3}\tilde{g}_{\lambda[\nu}\tilde{\nabla}_{\mu]}\tilde{T} 
\]
 
\medskip
\noindent
\textbf{Remark.} It should be noted that above the definitions and
decompositions are invariant under conformal
rescaling. Nevertheless, when raising indices or applying contractions
we have explicitly written out the metric to avoid ambiguity. In the
sequel, a frame formalism will be used throughout. This choice will
remove the ambiguity as the frame metric will always be
$\eta_{ij}$. Consistent with equation \eqref{frame:metric} the
metric $g_{\mu\nu}$ and its inverse will be used throughout for
raising and lowering tensorial indices.

\section{The extended conformal field equations with matter}
\label{Section:XCFE}

The idea of \emph{vacuum conformal Einstein field equations} expressed
in terms of the Levi-Civita connection $\nabla$ of a conformally
rescaled metric $g_{\mu\nu}$ and associated objects was originally
introduced in \cite{Fri81b,Fri81a,Fri83}. The generalisation of these
conformal equations to physical spacetimes containing matter was
discussed in \cite{Fri91}. More recently, a more general type of
vacuum conformal equations ---the \emph{extended conformal Einstein
field equations}--- expressed in terms of a Weyl connection
$\hat{\nabla}$ has been introduced ---see \cite{Fri95}. In this section we discuss how
these extended conformal field equations can be modified to discuss
spacetimes with matter.

\subsection{Frame formulation}

As in the previous section, let $e_k$ denote a frame which orthogonal
with respect to the metric $g_{\mu\nu}$ so that equation
\eqref{frame:metric} holds. In order to discuss the extended conformal
Einstein field equations, it will be convenient to depart slightly from
the point of view taken in the previous section and
regard,  for the moment,  the connection $\nabla$ only as a metric connection with
respect to $g_{\mu\nu}$ ---i.e. $\nabla_\lambda g_{\mu\nu}=0$.
Under this assumption, the connection $\nabla$ could have torsion, and thus it would not be a
Levi-Civita connection. The connection coefficients $\Gamma_i{}^k{}_j$
of $\nabla$ with respect to the frame $e_k$ are defined by the
relation
\[
\nabla_i e_j = \Gamma_i{}^k{}_j e_k.
\]
As a consequence of having a metric connection the connection
coefficients satisfy
\[
\Gamma_i{}^k{}_j \eta_{kl} + \Gamma_i{}^k{}_l \eta_{kj}=0.
\]
The torsion $\Sigma_i{}^k{}_j$ of the connection $\nabla$ is defined by
\[
\Sigma_i{}^k{}_j e_k \equiv \left(\Gamma_i{}^k{}_j - \Gamma_j{}^k{}_i  \right)e_k.
\]
If $\Sigma_i{}^k{}_j=0$ so that the connection $\nabla$ is the unique
Levi-Civita connection of $g_{\mu\nu}$, the connection coefficients
acquire the additional symmetry 
\[
\Gamma_i{}^k{}_j = \Gamma_j{}^k{}_i.
\]

\medskip
Now, given the connection coefficients $\Gamma_i{}^k{}_j$ of a metric
connection as above and a 1-form $f_\mu$, one can define a further
connection $\hat{\Gamma}_i{}^k{}_j$ using the relation
\begin{equation}
\hat{\Gamma}_i{}^k{}_j = \Gamma_i{}^k{}_j + S_{ij}{}^{kl}f_l,
\label{WeylToUnphysical}
\end{equation}
---cfr. \eqref{ChangeOfConnection:2}. Let
$\hat{\Sigma}_i{}^j{}_k$ denote the torsion of the connection $\hat{\nabla}$. It
follows directly that
\begin{equation}
\Sigma_i{}^k{}_j=\hat{\Sigma}_i{}^k{}_j. 
\label{ConformalInvariance:Torsion}
\end{equation}
so that $\hat{\nabla}$ will not be a Weyl connection unless
$\hat{\Sigma}_i{}^k{}_j=0$.

\medskip
In our subsequent discussion it will be convenient to distinguish
between the \emph{geometric curvature} $\hat{r}^k{}_{lij}$ ---i.e. the
expression of the curvature related to the
connection coefficients $\hat{\Gamma}_i{}^j{}_k$--- and the
\emph{algebraic curvature} $\hat{R}^k{}_{lij}$ ---i.e. the
decomposition of the curvature in terms of irreducible components. One
has that
\begin{subequations}
\begin{eqnarray*}
&& \hat{r}^k{}_{lij} \equiv e_i \left( \hat{\Gamma}\tensor{j}{k}{l} \right) - e_j \left( \hat{\Gamma}\tensor{i}{k}{l} \right) 
- \hat{\Gamma}\tensor{m}{k}{l} \left( \hat{\Gamma}\tensor{i}{m}{j} - \hat{\Gamma}\tensor{j}{m}{i}\right) 
+ \hat{\Gamma}\tensor{i}{k}{m}  \hat{\Gamma}\tensor{j}{m}{l} 
- \hat{\Gamma}\tensor{j}{k}{m}  \hat{\Gamma}\tensor{i}{m}{l} + \hat{\Sigma}_i{}^m{}_j \hat{\Gamma}_m{}^k{}_l, \\
&& \hat{R}^k{}_{lij} \equiv C^k{}_{lij} +2 \left( \delta^k{}_{[i} \hat{P}_{j]l} -  \delta^k{}_l \hat{P}_{[ij]} -  \eta_{l[i}\hat{P}_{j]}{}^k \right)
 = C^k{}_{lij} + 2 S_{l[i}{}^{km}\hat{P}_{j]m}
.
\end{eqnarray*}
\end{subequations}

For ease of the subsequent discussion we introduce the following
\emph{zero quantities}:
\begin{subequations}
\begin{eqnarray}
&&\hat{\Sigma}_i{}^l{}_j e_l\equiv \left( \hat{\Gamma}\tensor{i}{l}{j} - \hat{\Gamma}\tensor{j}{l}{i}\right) e_l -[e_i,e_j],\label{torsionfree}\\
&& \hat{\Xi}^k{}_{lij}\equiv \hat{r}^k{}_{lij} -\hat{R}^k{}_{lij}\label{Curvature}\\
&& \hat{\Delta}_{ij} \equiv \hat{\nabla}_{i} f_j - \hat{\nabla}_j f_i
-\hat{P}_{ij}+\hat{P}_{ji},  \label{ContractionCurvature}\\
\label{Cotton-York}
&& \hat{\Delta}_{lij} \equiv \hat{\nabla}_{i}\hat{P}_{jl} -
\hat{\nabla}_{j}\hat{P}_{il} - b_k C^k{}_{lij} - 
\tilde{Y}_{ijl}
\\
&& \hat{\Lambda}'_{lij}\equiv \hat{\nabla}_k C^k{}_{lij}  - 
 b_k C^k{}_{lij}
-\tilde{Y}_{ijl}.  \label{Bianchi} 
\end{eqnarray}
\end{subequations}

The interpretation of the zero quantities
 \eqref{torsionfree}-\eqref{Bianchi} is as follows:
 the zero quantity given by \eqref{torsionfree} measures the torsion
 of the connection $\hat{\nabla}$; that of \eqref{Curvature} relates
 the expression of the curvature of $\hat{\nabla}$ with its
 decomposition in terms of irreducible components; equation
 \eqref{ContractionCurvature} is contraction of
 \eqref{Curvature} over the first two indices. 
 It is included here for later convenience. 
 Equations \eqref{Cotton-York} and \eqref{Bianchi}
 measure the deviation from the fulfilment of the Bianchi identity. 

\medskip
The \emph{extended conformal Einstein field equations with matter} are
then given by
\begin{equation}
\hat{\Sigma}_i{}^k{}_j e_k=0, \quad \hat{\Xi}^k{}_{lij}=0, \quad \hat{\Delta}_{lij}=0,
\quad \hat{\Lambda}_{lij}=0. \label{XCFEFrame}
\end{equation}
These equations yield differential conditions for the frame
coefficients $e_i$, the spin coefficients $\hat{\Gamma}_i{}^j{}_k$,
the components of the 1-form $f_i$, the components of the Schouten
tensor $\hat{P}_{ij}$, and the Weyl tensor $C^k{}_{lij}$,
respectively.  The latter need to be complemented with the
energy-momentum conservation equation
\[
\tilde{\nabla}^i \tilde{T}_{ij}=0,
\]
whose particular details will depend on the matter model under
consideration. 

\medskip
Note that in equations \eqref{XCFEFrame}, the 1-form $b_k$ relating $\hat{\nabla}$ and $\tilde{\nabla}$ remains unspecified. In the sequel it will be convenient to introduce the
variables
\begin{subequations}
\begin{eqnarray}
&& d^k{}_{lij}\equiv \Theta^{-1}C^k{}_{lij}, \label{defining dklij}\\ 
&& d_i \equiv \Theta b_i = \nabla_i \Theta + \Theta f_i. \label{defining di}
\end{eqnarray}
\end{subequations}
In terms of the latter, the last two conformal field equations then read:
\begin{subequations}
\begin{eqnarray}
&&\hat{\Delta}_{lij} =\hat{\nabla}_{i}\hat{P}_{jl} - \hat{\nabla}_{j}\hat{P}_{il} -d_k d^k{}_{lij} -
\tilde{Y}_{ijl}=0, \label{Cotton-York2} \\
&& \Theta^{-1}\hat{\Lambda}'_{lij} = \hat{\Lambda}_{lij} =\hat{\nabla}_k d^k{}_{lij} - 
\Theta^{-1}\tilde{Y}_{ijl} - f_k d^k{}_{lij}=0. \label{Bianchi2} 
\end{eqnarray}
\end{subequations}
As in the case of $b_k$, the newly introduced function $\Theta$ and
the 1-form $d_k$ remain unspecified at this stage. They will latter be
fixed by the choice of a suitable conformal gauge.

\medskip
\noindent
\textbf{Remark 1.} If the extended conformal field equations
\eqref{XCFEFrame} are satisfied then the frame $e_k$ can be used to
construct a metric $g_{\mu\nu}$ via the relation
\eqref{frame:metric}. The connection coefficients
$\hat{\Gamma}_i{}^j{}_k$ give rise to a torsion-free connection, so that the
connection $\Gamma_i{}^j{}_k$ given by \eqref{WeylToUnphysical} is the
Levi-Civita connection of $g_{\mu\nu}$. Consequently, 
$\hat{\Gamma}_i{}^j{}_k$ defines a Weyl connection with conformal Weyl
tensor given by $C^k{}_{lij}$ and Schouten tensor
$\hat{P}_{ij}$. Showing that the solution so obtained implies a
solution to the Einstein field equations requires bringing into
consideration gauge conditions. This will be discussed together with
the propagation of the constraints in section
\ref{Section:PropagationOfConstraints}.

\medskip
\noindent
\textbf{Remark 2.} As a consequence of the transformation rules for
the Schouten tensor
\eqref{SchoutenPhysicalToWeyl1}-\eqref{SchoutenPhysicalToWeyl2}
and \eqref{SchoutenUnphysicalToWeyl1}-\eqref{SchoutenUnphysicalToWeyl2},
the zero quantities \eqref{torsionfree}-\eqref{Bianchi} involved in
the extended conformal field equations \eqref{XCFEFrame} transform
covariantly (i.e. homogeneously) under a change in the conformal
gauge. Thus, if they are satisfied in one gauge, then they are
satisfied in all gauges.

\subsection{Spinorial formulation}
In the sequel we will make use of a spinorial version of
the extended conformal field equations \eqref{XCFEFrame}. The use of this type 
of representation leads to simplifications, in particular, when
obtaining a reduced system of propagation equations. However, we will
switch to a frame representation whenever it is more convenient for
the discussion.

\medskip
The connection between the components of a tensor with respect to an
orthonormal basis and its spinorial counterpart is realised by the
constant Infeld-van der Waerden symbols.  In particular, let
$e_{AA'}$, $\hat{\nabla}_{AA'}$, $d_{AA'}$ 
denote, respectively,
the spinorial counterparts of $e_i$, $\hat{\nabla}_i$, $d_i$. Furthermore, let
$\Gamma_{AA'}{}^{BB'}{}_{CC'}$, $\hat{\Gamma}_{AA'}{}^{BB'}{}_{CC'}$
denote, respectively, the spinorial counterpart of the connection
coefficients $\Gamma_i{}^j{}_k$, $\hat{\Gamma}_i{}^j{}_k$. As the connection defined by $\Gamma_i{}^j{}_k$ is assumed to be metric, it follows that one can write
\[
\Gamma_{AA'}{}^{BB'}{}_{CC'} = \Gamma_{BB'}{}^A{}_C \epsilon_{C'}{}^{B'} + \bar{\Gamma}_{A'A}{}^{B'}{}_{C'}\epsilon_C{}^B, \quad \Gamma_{AA'BC}=\Gamma_{AA'(BC)}. 
\]
The symmetry condition on the last pair of indices of the spin
connection coefficient encodes the assumption of having a metric
connection. For the spin Weyl connection coefficients one has 
\[
\hat{\Gamma}_{AA'}{}^{BB'}{}_{CC'} = \hat{\Gamma}_{AA'}{}^B{}_C \epsilon_{C'}{}^{B'} + \bar{\hat{\Gamma}}_{A'A}{}^{B'}{}_{C'}\epsilon_C{}^B, 
\]
with 
\[
\hat{\Gamma}_{AA'}{}^B{}_C = \Gamma_{AA'}{}^B{}_C + \epsilon_{A}{}^{B}f_{CA'},
\quad \hat{\Gamma}_{AA'CB}=\hat{\Gamma}_{AA'(CB)}.
\]

\medskip 
Let $\hat{r}^{AA'}{}_{BB'CC'DD'}$ denote the spinorial counterpart of
the geometric curvature $\hat{r}^k{}_{lij}$. For future use we note
the Ricci identity:
\begin{equation}
\label{GeneralRicciIdentity}
\left( \hat{\nabla}_{CC'} \hat{\nabla}_{DD'} -
  \hat{\nabla}_{DD'}\hat{\nabla}_{CC'}   \right) \mu^{AA'} =
\hat{r}^{AA'}{}_{BB'CC'DD'} \mu ^{BB'} - \hat{\Sigma}_{CC'}{}^{EE'}{}_{DD'} \hat{\nabla}_{EE'}\mu^{AA'},
\end{equation}
valid for any spinor $\mu^{AA'}$ and where 
$\hat{\Sigma}_{CC'}{}^{EE'}{}_{DD'}$ is the spinorial counterpart of
the torsion. In our conventions, the geometric and algebraic curvature tensor of a general Weyl connection satisfy 
\begin{eqnarray*}
&&\hat{r}_{klij} = \hat{r}_{[kl]ij} + 2 \eta_{kl}\hat{\nabla}_{[i}f_{j]},   \\
&&\hat{R}_{klij} = \hat{R}_{[kl]ij} - 2 \eta_{kl}\hat{P}_{[ij]}.
\end{eqnarray*}
Their spinorial counterpart can be decomposed as
\begin{eqnarray*}
&&\hat{r}_{AA'BB'CC'DD'} = \epsilon_{A'B'} \hat{r}_{ABCC'DD'} + \epsilon_{AB} \bar{\hat{r}}_{A'B'CC'DD'}, \nonumber \\
&&\hat{R}_{AA'BB'CC'DD'} = \epsilon_{A'B'} \hat{R}_{ABCC'DD'} + \epsilon_{AB} \bar{\hat{R}}_{A'B'CC'DD'}
\end{eqnarray*}
where
\begin{eqnarray*}
&&\hat{r}_{ABCC'DD'}=\hat{r}_{(AB)CC'DD'} + \frac{1}{2}\epsilon_{AB} (\hat{\nabla}_{CC'}f_{DD'}-\hat{\nabla}_{DD'}f_{CC'}), \nonumber \\
&&\hat{R}_{ABCC'DD'}=\hat{R}_{(AB)CC'DD'} - \frac{1}{2}\epsilon_{AB} (\hat{P}_{CC'DD'}-\hat{P}_{DD'CC'}).
\end{eqnarray*}
In terms of the reduced spin coefficients $\hat{\Gamma}_{AA'BC}$ one
writes the geometric curvature (assuming that the torsion vanishes) as
\begin{eqnarray*}
&& r^C{}_{DAA'BB'} =e_{AA'} \left( \hat{\Gamma}\tensor{BB'}{C}{D} \right) - e_{BB'} \left( \hat{\Gamma}\tensor{AA'}{C}{D} \right)- \hat{\Gamma}\tensor{AA'}{F}{B} \hat{\Gamma}\tensor{FB'}{C}{D} + \hat{\Gamma}\tensor{BB'}{F}{A} \hat{\Gamma}\tensor{FA'}{C}{D}
\nonumber \\
&& \hspace{1cm}- \overline{\hat{\Gamma}}_{AA'}{}^{F'}{}_{B'} \hat{\Gamma}_{BF'}{}^{C}{}_{D} +\overline{\hat{\Gamma}}{}_{BB'}{}^{F'}{}_{A'} \hat{\Gamma}\tensor{AF'}{C}{D} + \hat{\Gamma}\tensor{AA'}{C}{F}\hat{\Gamma}\tensor{BB'}{F}{D} - \hat{\Gamma}\tensor{BB'}{C}{F}\hat{\Gamma}\tensor{AA'}{F}{D}.
\end{eqnarray*}

\subsubsection{The uncontracted spinorial conformal field equations}
The spinorial version of the zero quantities
\eqref{torsionfree}-\eqref{Bianchi} is given by
\begin{subequations}
\begin{eqnarray}
&& \hat{\Sigma}_{AA'}{}^{PP'}{}_{BB'}e_{PP'}\equiv [e_{AA'},e_{BB'}]-\hat{\Gamma}\tensor{AA'}{C}{B} e_{CB'} - \overline{\hat{\Gamma}}\tensor{AA'}{C'}{B'}e_{BC'}  \nonumber \\
&& \hspace{4cm}+ 
\hat{\Gamma}\tensor{BB'}{C}{A} e_{CA'} +
\overline{\hat{\Gamma}}\tensor{BB'}{C'}{A'}e_{AC'} , \label{Xcfe1}\\
&& \hat{\Xi}_{ABCC'DD'} \equiv \hat{r}_{ABCC'DD'} -\hat{R}_{ABCC'DD'},\label{Xcfe2} \\
&& \hat{\Delta}_{CC'AA'BB'}\equiv \hat{\nabla}_{AA'}\hat{P}_{BB'CC'} - \hat{\nabla}_{BB'}\hat{P}_{AA'CC'} \nonumber\\
&& \hspace{2.5cm}-d^{PP'} d_{PP'AA'BB'CC'} 
 -\tilde{Y}_{AA'BB'CC'}, \label{Xcfe3}\\
&& \hat{\Lambda}_{CC'AA'BB'} \equiv \hat{\nabla}^{PP'}
d_{PP'CC'AA'BB'} - f^{PP'} d_{PP'CC'AA'BB'}- \Theta^{-1} \tilde{Y}_{AA'BB'CC'} \label{Xcfe4}
\end{eqnarray}
\end{subequations}
In terms of these spinorial zero quantities, the extended conformal
field equations are given by
\begin{equation}
\hat{\Sigma}_{AA'}{}^{PP'}{}_{BB'}e_{PP'}=0, \quad \hat{\Delta}_{ABCC'DD'}=0,
\quad \hat{\Delta}_{AA'BB'CC'}=0, \quad \hat{\Lambda}_{AA'BB'CC'}=0. \label{SpinorialXCFE}
\end{equation}

\subsubsection{The contracted spinorial conformal field equations}
Equations \eqref{Xcfe1}-\eqref{Xcfe3} are antisymmetric upon
interchange of a pair of indices. This structural
property will be used to obtain a contracted version of the equations
which will be systematically used in the sequel. The associated zero
quantities are given by
\begin{subequations}
\begin{eqnarray}
&& \tfrac{1}{2} \hat{\Sigma}_{(A|Q|}{}^{PP'}{}_{B)}{}^Q e_{PP'} \equiv
\hat{\nabla}_{(A|Q'|} e_{B)}{}^{Q'} - \hat{\Gamma}_{(A|Q'|}{}^P{}_{B)}
e_P{}^{Q'} - \bar{\hat{\Gamma}}_{Q'(A}{}^{P'Q'} e_{B)P'}, \\
&& \tfrac{1}{2}\hat{\Xi}_{(AB)CQ'D}{}^{Q'} \equiv
\tfrac{1}{2}\left(\hat{r}_{ABCQ'D}{}^{Q'} - \hat{R}_{ABCQ'D}{}^{Q'}
\right), \\
&& \tfrac{1}{2} \hat{\Delta}_{CC'(A|Q'|B)}{}^{Q'} =
\hat{\nabla}_{(A|Q'|}\hat{P}_{B)}{}^{Q'}  +d^Q{}_{C'} \phi_{ABCQ} -\tilde{Y}_{ABCC'},
\end{eqnarray}
\end{subequations}
and their complex conjugate versions. Above the symmetries of the spinorial counterparts of $d_{klij}$ and
$\tilde{Y}_{ijk}$ have been exploited by writing
\begin{eqnarray*}
&& d_{AA'BB'CC'DD'} = \phi_{ABCD}\epsilon_{A'B'} \epsilon_{C'D'} + \overline{\phi}_{A'B'C'D'}\epsilon_{AB} \epsilon_{CD}, \\
&& \tilde{Y}_{AA'BB'CC'} = \tilde{Y}_{ABCC'}\epsilon_{A'B'} + \overline{\tilde{Y}}_{A'B'C'C}\epsilon_{AB},
\end{eqnarray*}
with
\[
\phi_{ABCD}=\phi_{(ABCD)}, \quad \tilde{Y}_{ABCC'}=\tilde{Y}_{(ABC)C'}.
\]
Using the latter formulae equation \eqref{Xcfe4} reduces to its more
usual form:
\[
\hat{\Lambda}_{A'ABC}\equiv \hat{\nabla}^Q{}_{A'}\phi_{ABCQ} - f^Q{}_{A'}
\phi_{ABCQ} -\Theta^{-1}\tilde{Y}_{ABCA'}.
\]

\medskip
The extended conformal field equations can be expressed in terms of
these contracted zero quantities as:
\begin{equation}
\hat{\Sigma}_{(A|Q|}{}^{PP'}{}_{B)}{}^Q e_{PP'}=0, \quad
\hat{\Xi}_{(AB)CQ'D}{}^{Q'}=0, \quad
\hat{\Delta}_{CC'(A|Q'|B)}{}^{Q'}, \quad \hat{\Lambda}_{A'ABC}=0. \label{ContractedSpinorialXCFE}
\end{equation}
It is important to remark that the contracted equations
\eqref{ContractedSpinorialXCFE} are fully equivalent to \eqref{SpinorialXCFE}.

\section{Conformal curves and generalised conformal Gaussian gauge systems}
\label{Section:ConformalCurves}

The advantage of considering extended conformal equations in terms of
Weyl connections is that they allow to consider gauge systems based on
conformally invariant objects. An example of these gauge systems are
the \emph{conformal Gaussian systems} introduced in
\cite{Fri95,Fri03c}.  These gauge systems are based on
conformal geodesics. Conformal Gaussian systems are of great utility
in the discussion of evolution problems for the conformal field
equations as they provide a canonical conformal factor as well as structural
simplifications in the form of the evolution equations.

\medskip
It was shown in \cite{Fri95,Fri03c} that for vacuum 
spacetimes the conformal factor is quadratic in the conformal time and can be 
read of from the initial data of the evolution system. Accordingly,
the location of the conformal boundary is know \emph{a priori}. 
This predetermined character of the conformal factor hinges crucially on the fact that the
physical spacetime is vacuum. In the sequel we show that the conformal
geodesic equations in the presence of matter can be modified in such a
way that one has again a conformal factor known \emph{a priori}.

\subsection{A class of conformal curves}
Let $I\in \Real$ be an open interval. We will consider a class of
\emph{conformal curves}, $x^\mu(\tau)$, whose tangent vector
$v^\mu\equiv \dot{x}^\mu$ is coupled to a 1-form $b_\nu$ 
via the equations
\begin{subequations}
\begin{eqnarray}
&& \tilde{\nabla}_v v^\rho + v^\mu v^\nu S\tensor{\mu\nu}{\rho\lambda}{} b_\lambda = 0 , \label{Gcg1}\\
&&\tilde{\nabla}_v b_\nu - \half v^\mu S\tensor{\mu\nu}{\rho\lambda}{}b_\rho b_\lambda = \tilde{H}_{\mu\nu}v^\mu . \label{Gcg2}
\end{eqnarray}
\end{subequations}
where $\tilde{H}_{\mu\nu} $ transforms under $\nabla = \tilde{\nabla} + S(\Upsilon) $ as
\begin{equation} 
\label{generalised Schouten}
\tilde{H}_{\mu\nu} - H_{\mu\nu} 
= \tilde{\nabla}_\mu \Upsilon_\nu - \half  S \tensor{\mu\nu}{\rho\lambda}{}\Upsilon_\rho \Upsilon_\lambda 
= \nabla_\mu \Upsilon_\nu + \half  S \tensor{\mu\nu}{\rho\lambda}{}\Upsilon_\rho \Upsilon_\lambda .
\end{equation}
Equations \eqref{Gcg1}-\eqref{Gcg2} will be
supplemented with a frame propagation equation via
\begin{equation} 
\label{Gcg3}
  \tilde{\nabla}_v e^\rho_k + v^\mu e^\nu_k S\tensor{\mu\nu}{\rho\lambda}{}b_\lambda = 0 
\end{equation} 
In a slight abuse of terminology, we will call a triple
$(v^\mu,b_\nu, e^\mu_k )$ solving equations \eqref{Gcg1}-\eqref{Gcg2}
and \eqref{Gcg3} a \emph{conformal curve}, since these curves exhibit
the conformally invariant behaviour described in the following lemma.

\begin{lemma}
\label{Lemma:GCGG}
Let $(x^\mu (\tau), b_\nu (\tau), e^\mu_k (\tau) )$ be a conformal curve. Then
$(x^\mu (\tau), (b_\nu - h_\nu) (\tau), e^\mu_k (\tau) )$ satisfies \eqref{Gcg1},
\eqref{Gcg2} and \eqref{Gcg3} expressed in the connection $\check{\nabla} =
\tilde{\nabla} + S(h) $. In particular, in terms of the Weyl connection
given by $\hat{\nabla} = \tilde{\nabla} + S(b) $ the conformal curve
equations take the form
\begin{equation}
\label{GCGG:tensor}
  \hat{\nabla}_v v^\mu=0 , \quad \quad\hat{H}_{\mu\nu}v^\mu=0 , \quad \quad \hat{\nabla}_v e^\mu_k=0.
\end{equation}
Moreover, conformal curves are preserved as point sets under
reparametrisations of $\tau$ by fractional linear transformations.
\end{lemma}

\medskip
\noindent
\textbf{Remark.} Comparing \eqref{SchoutenPhysicalToWeyl1}-\eqref{SchoutenUnphysicalToWeyl2} and \eqref{generalised
Schouten} we can see that
\begin{eqnarray*}
&& \tilde{J}_{\mu\nu} \equiv \tilde{P}_{\mu\nu} - \tilde{H}_{\mu\nu}
\\
&&\phantom{\tilde{J}_{\mu\nu}}=\hat{P}_{\mu\nu} - \hat{H}_{\mu\nu} \\
&&\phantom{\tilde{J}_{\mu\nu}}= P_{\mu\nu} - H_{\mu\nu} .
\end{eqnarray*}
Thus, $\hat{H}_{\mu\nu}v^\mu=0 $ implies 
\[
\hat{P}_{\mu\nu}v^\mu=\tilde{J}_{\mu\nu} v^\mu.
\]


\bigskip
A vector frame satisfying \eqref{Gcg3} is called \emph{Weyl
propagated}. It is noted that the velocity can be chosen as one of the
frame vectors due to \eqref{Gcg1}. Suppose along a conformal curve we
define 
\[
\Theta(\tau) \equiv \vert \tilde{g}(v,v) \vert^{-1/2}.
\]
Then $\Theta(\tau)$ satisfies
\begin{equation}
\label{Theta evolution}
\dot{\Theta} = \Theta \langle b , v \rangle
\end{equation}
Letting $g_{\mu\nu}$ be given as in \eqref{rescaled:metric}, one finds
that $g(v,v)=1$. In fact, for a frame $e^\mu_k $ the frame metric 
\[
\eta_{jk} \equiv g(e_j, e_k)
\]
 is constant along the curve. Hence a $g$-orthonormal frame evolves
into a $g$-orthonormal frame along the curve. Following an analogous
discussion for conformal geodesics given in \cite{Fri03c} one
differentiates \eqref{Theta evolution} twice along the curves and
substitutes \eqref{Gcg1} and \eqref{Gcg2} to obtain
\begin{equation}
\label{d3Theta}
\dddot{\Theta}= \left( \tilde{\nabla}_v (\tilde{H}(v,v)) + \tilde{H}(v,b) \tilde{g}(v,v)
+ \langle b,v \rangle \tilde{H}(v,v)
\right) \Theta. 
\end{equation}
Note that for $\tilde{H}_{\mu\nu} = \lambda \tilde{g}_{\mu\nu} $ the
right hand side vanishes exactly. Thus, the following result holds:

\begin{lemma}
\label{Lemma:ConformalFactor}
Suppose that $(v^\mu(\tau),b_\nu(\tau),e_k(\tau))$ is a solution to the
conformal curve equations \eqref{Gcg1}, \eqref{Gcg2}
and \eqref{Gcg3} with respect to the metric $\tilde{g}_{\mu\nu}$ such
that $x(\tau)$ is a timelike curve in $\tilde{\mathcal{M}}$ defined on some open
interval $I$. If $\tilde{g}_{\mu\nu}$ satisfies the Einstein equations
with matter then:
\begin{itemize}
\item[i)]
\[
\tilde{H}_{\mu\nu} = \lambda \tilde{g}_{\mu\nu} \Leftrightarrow
\tilde{J}_{\mu\nu} = \frac{1}{2} \left( \tilde{T}_{\mu\nu} -
  \frac{1}{3}\tilde{T}_{\rho\sigma}\tilde{g}^{\rho\sigma}\tilde{g}_{\mu\nu}
\right);
\]

\item[ii)] the conformal factor $\Theta$ is given for $\tau\in I$ by
\begin{equation}
\label{canonical:vacuum}
\Theta = \Theta_* + \dot{\Theta}_*(\tau-\tau_*) + \ddot{\Theta}_* (\tau-\tau_*)^2,
\end{equation}
where a quantity with a subscript $*$ is constant along $x(\tau)$.
\end{itemize}
\end{lemma}

\medskip

For vacuum spacetimes one has $\tilde{J}_{\mu\nu} = 0$ and one
recovers known results for conformal geodesics ---see
e.g. \cite{Fri95,Fri03c}. In the presence of matter the conformal
curves curves given by equations \eqref{Gcg1}-\eqref{Gcg2} are no
longer conformal geodesics. However, our choice of $\tilde{H}_{\mu\nu}$
gives the same behaviour for the \emph{canonical} conformal factor as
in the vacuum case. We will also require the following result:

\begin{lemma}
\label{Lemma:d}
Suppose that $(v^\mu(\tau),b_\nu(\tau),e_k(\tau))$ is a conformal
curve  as in Lemma
\ref{Lemma:ConformalFactor}. Let $\tilde{g}_{\mu\nu}$ satisfy the
Einstein field equations with matter and $\tilde{H}_{\mu\nu} = \lambda
\tilde{g}_{\mu\nu}$. If $v^\mu =e^\mu_0$ at $\tau=\tau_0\in I$, then
\[
b_k(\tau) = b_\mu e^\mu_k = \Theta^{-1} \left( \dot{\Theta}, d_{a*}\right)
\]
for $\tau\in I$, where $d_{a*} = \Theta b_a(\tau_0)$, for $a=1,\;2,\;3$.
\end{lemma}

\noindent
The proof of this result is a calculation analogous to the one
described in the proof of Lemma 3.2 in \cite{Fri95}

\subsection{Jacobi fields for conformal curves}

Suppose we are given a congruence of conformal curves with velocity
$v$. The separation vector $\eta$ satisfies $[v,\eta] = 0 $ along each
conformal curve and will be referred to as a Jacobi field.  Recall
that Weyl connections are torsion-free, so that $[v,\eta] -
\hat{\nabla}_v \eta + \hat{\nabla}_\eta v = 0$. Thus we get an
evolution equation for $\eta$
\[
\partial_\tau\eta_k = \hat{\nabla}_v \eta^k = \eta^j (\hat{\nabla}_{j} v^{k} )
\]

%
%

When the Jacobi field becomes tangent to the curve at a point $p$ we
say that $p$ is a conjugate point. These points are of interest to us
for the following reason. If we create Gaussian coordinate system by
dragging spatial coordinates along the congruence beyond a conjugate
point then these are ill-defined. For this reason we measure $\vert
\eta - g(v,\eta)v \vert^2= (\eta^{ab} \eta_a \eta_b) $ with $a, b=1,2,3$. As long as this quantity does not vanish our
coordinate system will be well-defined.

\subsection{Generalised conformal Gaussian systems}
\label{Section:DefinitionGauge}

In analogy to the way conformal geodesics have been used in
\cite{Fri95,Fri98a,LueVal09a,LueVal09b}, to construct conformal
Gaussian gauge systems, here we will use the conformal curves solving
equations \eqref{Gcg1}-\eqref{Gcg2} and \eqref{Gcg3} to construct what
we will call \emph{generalised conformal Gaussian systems}.

\medskip
Let $\tilde{\mathcal{S}}$ be a space-like hypersurface in the spacetime
$(\tilde{\mathcal{M}},\tilde{g}_{\mu\nu})$. On $\tilde{\mathcal{S}}$ we choose
an initial conformal factor $\Theta_*>0$, a frame field $e_{k*}$, and
a 1-form $b_*$ such that
\begin{equation}
\label{DataGcg1}
\Theta^2_*\tilde{g}(e_{i*},e_{j*})=\eta_{ij},
\end{equation}
and $e_{0*}$ is
orthogonal to $\tilde{\mathcal{S}}$.  For fixed $\tilde{H}_{\mu\nu}$ and given
$x_*\in\tilde{\mathcal{S}}$ there exists a unique conformal curve
$(x^\mu(\tau),b_\nu(\tau))$, which for $\tau=0$ passes through $x_*$
and which satisfies the initial conditions 
\begin{equation}
\dot{x}^\mu=e_{0*}^\mu, \quad b_\nu = b_{\nu*}.
\label{DataGcg2}
\end{equation}
If all data are
smooth, then in some neighbourhood $\mathcal{U}\in\tilde{\mathcal{S}}$ 
these curves define a smooth caustic free congruence
covering $\mathcal{U}$. Furthermore, $b_\nu$ defines a smooth
1-form on $\mathcal{U}$ which allows to construct a Weyl connection
$\hat{\nabla}$. A smooth frame field $e^\mu_k$ and the related
conformal factor $\Theta$ are obtained in $\mathcal{U}$ by solving the
propagation equations \eqref{Gcg3} and \eqref{Theta evolution} for given
initial data 
\begin{equation}
e^\mu_k=e^\mu_{k*}, \quad \Theta=\Theta_*
\label{DataGcg3}
\end{equation}
on $\tilde{\mathcal{S}}$. Then $e_0^\mu(\tau) = \dot{x}^\mu(\tau)$ on
$\mathcal{U}$ and we define $$\hat{\chi}_i{}^j{} \equiv \hat{\nabla}_i
v^j= \hat{\Gamma}_i{}^j{}_0 .$$ The frame one obtains from solving the
propagation equation is orthonormal for the metric
$g_{\mu\nu}=\Theta^2\tilde{g}_{\mu\nu}$, while $
\hat{\Gamma}_0{}^j{}_k =0$.  Dragging along local coordinates $x^a$ on
$\tilde{\mathcal{S}}$ with the congruence and setting $x^0=\tau$, one
obtains a coordinate system. A coordinate system, a frame field and a
conformal factor constructed with the above procedure will be known as
a \emph{generalised conformal Gaussian system}.

We will follow the setup used in \cite{LueVal09a, LueVal09b}, where we
use a global frame field $c_\es $ ($\es=0, 1, 2, 3$) constructed from
the coordinate vectors in such a way that $c_0 = \partial_\tau$ and
$c_\er $ ($\er= 1, 2, 3$) is a constant linear combination of the
spatial coordinate vectors. The frame $e_k$ then is written in terms
of its expansion $e_k = e_k^\es c_\es $. The same is done for the
spinorial version.

\medskip
\noindent
\textbf{Remark.} 
Above we have set up a generalised conformal Gaussian system in the
physical spacetime
$(\tilde{\mathcal{M}},\tilde{g}_{\mu\nu})$. However, 
due to the conformal invariance of conformal curves proven in
Lemma\ref{Lemma:GCGG} the same gauge can also be constructed starting
from an spacelike hypersurface $\check{\mathcal{S}}$ in a conformally
related spacetime $(\check{\mathcal{M}},\check{g}_{\mu\nu})$ such that
$\tilde{\mathcal{S}} \subset \check{\mathcal{S}},$ $\tilde{\mathcal{M}}
\subset \check{\mathcal{S}}$. The initial data will be related in the
obvious way, with $\Theta_*$ and $b_*$ changing accordingly while the
frame remain the same.  We will make us of this fact later on when
constructing a generalised conformal Gaussian system from
hyperboloidal data given on $\mathcal{S}$ in the unphysical spacetime
$(\mathcal{M},g_{\mu\nu})$.

\medskip
Due to Lemma\ref{Lemma:GCGG} a generalised conformal Gaussian system is
characterised on $\mathcal{U}$ by the explicit conditions
\begin{equation}
\label{GaussianGauge:Tensorial}
v^\mu = \partial_\tau, \quad \hat{\Gamma}_0{}^j{}_k =0, \quad \hat{P}_{0k}=\tilde{J}_{0k}. 
\end{equation}
Setting 
\[
\tilde{J}_{\mu\nu} = \frac{1}{2} \left( \tilde{T}_{\mu\nu} -
  \frac{1}{3}\tilde{T}_{\rho\sigma}\tilde{g}^{\rho\sigma}\tilde{g}_{\mu\nu}
\right)
\]
one has, by virtue of Lemma \ref{Lemma:ConformalFactor}, that the
solution of the evolution of the conformal factor is known \emph{a
priori}. In the sequel \emph{we will only consider trace-free
matter} (the Maxwell field). Consequently,
\[
\tilde{J}_{\mu\nu} = \frac{1}{2} \tilde{T}_{\mu\nu}, \quad \tilde{\nabla}^\mu\tilde{J}_{\mu\nu} =0.
\]
The latter implies $\nabla^\mu J_{\mu\nu} =0 $ if one defines
$J_{\mu\nu}\equiv \Theta^{-2}\tilde{J}_{\mu\nu} $. 

\section{Hyperbolic reductions of the conformal field equations}
\label{Section:HyperbolicReductions}

In this section we discuss how to extract a symmetric hyperbolic
system of propagation equations. For this, we resort to a space-spinor
formalism ---see e.g. \cite{Som80}--- based on the spinorial counterpart, $\tau^{AA'}$, of a
timelike vector $\tau^\mu$ in the conformally rescaled
spacetime $(\mathcal{M},g_{\mu\nu})$. More precisely, the vector
$\tau^\mu$ will be taken to be parallel to the tangent vector $v^\mu$
to the conformal curves described in Section
\ref{Section:ConformalCurves}. The normalisation condition $\tau_\mu
\tau^\mu=2$ will be used.

\medskip
The reduced symmetric hyperbolic system
of evolution equations is to be deduced from the following
contractions of the conformal field equations
\begin{equation}
\tau^{AA'}\hat{\Sigma}_{AA'}{}^{PP'}{}_{BB'}e_{PP'}=0,  \quad
\tau^{CC'}\hat{\Xi}_{ABCC'DD'}=0, \quad
\tau^{AA'}\hat{\Delta}_{AA'BB'CC'}=0,  \label{ReducedXCFEa}
\end{equation}
together with
\begin{equation}
\tau_{(A}{}^{A'}\hat{\Lambda}_{|A'|B)CD} =0. \label{ReducedXCFEb}
\end{equation}

\subsection{The space spinor formalism in brief}
In what follows, we will consider spin dyads $\{\delta_A\}$ for which
the spinor $\tau^{AA'}$ admits the decomposition
\[
\tau^{AA'} = \epsilon_0{}^A \epsilon_{0'}{}^{A'} + \epsilon_1{}^A \epsilon_{1'}{}^{A'}. 
\]
In particular, one has that
\begin{equation}
\tau_{AA'} \tau^{BA'} = \epsilon_A{}^B.
\label{TauTauEpsilon}
\end{equation}

Using the spinor $\tau^{AA'}$, the gauge conditions
\eqref{GaussianGauge:Tensorial} can be rewritten as
\begin{equation}
\label{GaussianGauge:Spinorial}
\tau^{AA'} e_{AA'} = \sqrt{2} \partial_\tau, \quad \tau^{AA'}\hat{\Gamma}_{AA'}{}^B{}_C=0, \quad \tau^{AA'} \hat{P}_{AA'BB'}= \tilde{J}_{BB'}. 
\end{equation} 
where $\tilde{J}_{BB'} $ denotes the spinorial counterpart of $\tilde{J}_{ij}\tau^i$.

\medskip
The spinor $\tau^{AA'}$ can also be used to obtain an unprimed
version of the spinorial Weyl connection covariant derivative
$\hat{\nabla}_{AA'}$. More precisely, one has that: 
\[
\hat{\nabla}_{AB}\equiv \tau_B{}^{A'} \hat{\nabla}_{AA'}.
\]
The latter, in turn, can be decomposed in its irreducible parts:
\begin{equation}
\hat{\nabla}_{AB} = \tfrac{1}{2} \epsilon_{AB} \hat{\mathcal{P}} + \hat{\mathcal{D}}_{AB},
\label{SpaceSpinorNablaDecomposition}
\end{equation}
where 
\[
\hat{\mathcal{P}}\equiv \tau^{AA'} \hat{\nabla}_{AA'}, \quad
\hat{\mathcal{D}}_{AB}\equiv \tau_{(B}{}^{A'} \hat{\nabla}_{A)A'}.
\]
The differential operator $\hat{\mathcal{D}}_{AB}$ is the so-called
\emph{Sen connection} of $\hat{\nabla}$ relative to the vector field
$\tau^{AA'}$.

\subsection{Hyperbolic reduction of a first  model equation}
\label{Section:ModelEquation}
The Procedure and subtleties of deriving hyperbolic equations from the
extended conformal equations \eqref{SpinorialXCFE} will be illustrated
with a model equation. 

\medskip
Let $M_{i{\cal{K}}}$ and $N_{ij{\cal{K}}} = N_{[ij]{\cal{K}}}$ be two tensorial quantities, where $ {\cal{K}}$ stands for any set of tensor or bundle indices, satisfying the equation
\begin{equation}
\hat{\nabla}_i M_{j{\cal{K}}} - \hat{\nabla}_j M_{i{\cal{K}}} =
N_{ij{\cal{K}}}. \label{ModelEqn}
\end{equation}
Two derive an evolution equation we contract with $\tau^i$:
\[
\hat{\nabla}_\tau M_{j{\cal{K}}} - \hat{\nabla}_j
(M_{i{\cal{K}}}\tau^i ) =N_{ij {\cal{K}}} - M_{i {\cal{K}}}
(\hat{\nabla}_j \tau^i ) 
\]
from where it follows that
\[
\sqrt{2}\partial_\tau M_{j {\cal{K}}} - \hat{\nabla}_j (M_{i {\cal{K}}}\tau^i) =N^{'}_{ij {\cal{K}}} \equiv N_{ij {\cal{K}}} + \tau^i (\hat{\Gamma}\tensor{i}{l}{j} M_{l {\cal{K}}} +\hat{\Gamma}\tensor{i}{ {\cal{L}}}{ {\cal{K}}}M_{j {\cal{L}}} )- M_{i {\cal{K}}} (\hat{\nabla}_j \tau^i ) .
\]
Note that in our setup the connection coefficients in the expression vanish due to our gauge
choice. However, the following analysis is valid without this condition and it should be 
observed that these terms have a polynomial form.
We note that $M_{0 {\cal{K}}}$ appears inside the second
term. However, we can not obtain an evolution equation for $M_{0
  {\cal{K}}}$ from equation \eqref{ModelEqn} since setting $j=0$ in
order to get the evolution equation for $M_{0 {\cal{K}}}$ makes both
sides reduce trivially to zero due to the skew symmetry in $ij$. 
Instead, $M_{0 {\cal{K}}}$  must be determined from the symmetries of
$M_{i{\cal{K}}}$ (if any) or if not, then it must be regarded as free
data. 

\medskip
The subsequent discussion of the properties of the model equation
\eqref{ModelEqn} will be carried out in the spinor formalism. From the
spinorial version of \eqref{ModelEqn} 
\[
\hat{\nabla}_{AA'} M_{BB'{\cal{K}}} - \hat{\nabla}_{BB'} M_{AA'{\cal{K}}} =
N^{'}_{AA'BB'{\cal{K}}}. \label{ModelEqn}
\]
one obtains the contracted versions
\begin{subequations}
\begin{eqnarray}
&& \hat{\nabla}_{(A | P' |} M_{B)}{}^{P'}{}_{\cal{K}}  = \tfrac{1}{2}
N^{'}_{(A|P'|B)}{}^{P'}{}_{\mathcal{K}} , \label{ModelContracted1}
 \\
&& \hat{\nabla}_{P(A'} M_{B')}{}^{P'}{}_{\mathcal{K}}    = \tfrac{1}{2}
N^{'}_{P(A'}{}^{P'}{}_{B')\mathcal{K}}.  \label{ModelContracted2}
\end{eqnarray}
\end{subequations}
In order to change to space spinor components it is observed that 
one has to contract with $\tau\tensor{A}{A'}{}$ inside the derivative. One uses
that $M_{BB'{\cal{K}}} = - M_{BP{\cal{K}}}\tau\tensor{}{P}{B'} $ so
that
\begin{eqnarray*}
&&\tau\tensor{P}{A'}{}\tau\tensor{Q}{B'}{}\hat{\nabla}_{AA'} M_{BB' {\cal{K}}} = \tau\tensor{P}{A'}{}\tau\tensor{Q}{B'}{} \left( - \tau\tensor{}{R}{B'} \hat{\nabla}_{AA'} M_{BR {\cal{K}}} - M_{BR{\cal{K}}} \hat{\nabla}_{AA'} \tau\tensor{}{R}{B'} \right)\\
&& \phantom{\tau\tensor{P}{A'}{}\tau\tensor{Q}{B'}{}\hat{\nabla}_{(AA'} M_{BB' {\cal{K}}}}= \hat{\nabla}_{AP} M_{BQ {\cal{K}}} - M_{BR{\cal{K}}} \, \hat{\chi}\tensor{AP}{R}{Q}.
\end{eqnarray*}

Thus, from \eqref{ModelContracted1}-\eqref{ModelContracted2} one obtains
\begin{eqnarray*}
&& \hat{\nabla}_{(A \vert P \vert} M\tensor{ B)}{P}{ {\cal{K}}}   =
\tfrac{1}{2} N^{''}{}\tensor{(A \vert P \vert B)}{P}{ {\cal{K}}} 
\equiv \tfrac{1}{2} N^{'}{}\tensor{(A \vert P \vert B)}{P}{ {\cal{K}}} - M\tensor{(A}{R}{\vert {\cal{K}}\vert } \, \hat{\chi}\tensor{B)PR}{P}{} \\
&& \hat{\nabla}_{A(P} M\tensor{}{A}{Q) {\cal{K}}}  = 
\tfrac{1}{2} N^{''}{}\tensor{A(P}{A}{Q) {\cal{K}}}  
\equiv \tfrac{1}{2} N^{'}{}\tensor{A(P}{A}{Q) {\cal{K}}} + M\tensor{}{A}{R{\cal{K}}} \, \hat{\chi}\tensor{A(P}{R}{Q)}.
\end{eqnarray*}
Using the decomposition \eqref{SpaceSpinorNablaDecomposition} for
$\hat{\nabla}_{AB}$ and writing $ M_{ AB {\cal{K}}}$as
\[  
M_{ AB {\cal{K}}} = \half \epsilon_{AP} m_{{\cal{K}}} + m_{ AB
  {\cal{K}}}.
\]
one obtains:
\begin{eqnarray*}
&&  - \tfrac{1}{2} \hat{{\cal{P}}} m_{ (AB) {\cal{K}}} + \tfrac{1}{2}
\hat{{\cal{D}}}_{ AB } m_{ {\cal{K}}} + \hat{{\cal{D}}}_{P(A }
m_{B)}{}^P{}_{\cal{K}}    = \tfrac{1}{2}  N^{''}{}\tensor{(A \vert P \vert B)}{P}{ {\cal{K}}}, \\
&&\tfrac{1}{2}\hat{{\cal{P}}}  m_{(PQ) {\cal{K}}} - \tfrac{1}{2}  \hat{{\cal{D}}}_{ PQ } m_{{\cal{K}}}+ \hat{{\cal{D}}}_{A(P} m^A{}_{Q)\cal{K}}     = \tfrac{1}{2} N^{''}{}\tensor{A(P}{A}{Q) {\cal{K}}}. 
\end{eqnarray*}
Making linear combinations of the latter equations one finally arrives at:
\begin{subequations}
\begin{eqnarray}
&&\hat{{\cal{P}}}  m_{AB {\cal{K}}} - \hat{{\cal{D}}}_{ AB } m_{  {\cal{K}}} 
= E^{[M]}_{AB {\cal{K}}} 
\equiv \tfrac{1}{2} \left(N^{''}{}\tensor{P(A}{P}{B) {\cal{K}}} 
-  N^{''}{}\tensor{(A \vert P \vert B)}{P}{ {\cal{K}}} \right), \label{ModelEvolution}\\
&& \hat{{\cal{D}}}_{P(A} m^P{}_{B)\mathcal{K}} 
= C^{[M]}_{AB {\cal{K}}} 
\equiv \tfrac{1}{2}\left( N^{''}{}\tensor{P(A}{P}{B) {\cal{K}}}+  N^{''}{}\tensor{(A
    \vert P \vert B)}{P}{ {\cal{K}}}\right). \label{ModelConstraint}
\end{eqnarray}
\end{subequations}
The terms $ E^{[M]}_{AB {\cal{K}}} $ and $ C^{[M]}_{AB {\cal{K}}} $
introduced on the right hand side are formed from the original term $
N_{AB {\cal{K}}} $ and connection coefficients. It will be seen that
in their explicit form they are polynomial in the variables of our
system.  We will write the original variable, here $M$, in square
brackets and the variables $E $ and $C$ are used to indicate whether
the term is for the evolution or the constraint equation.

\medskip
Equation \eqref{ModelEvolution} is an evolution equation for the
spinorial components $m_{AB\mathcal{K}}$. In what concerns the
``timelike'' components $m_{\mathcal{K}}$ (i.e $M_{0\mathcal{K}}$) one
will have two possible situations:
\begin{itemize}
\item[(i)] there exists an external equation that relates
$m_{\mathcal{K}}$ and $m_{AB\mathcal{K}}$ and possibly some other
variables to each other. Then \eqref{ModelEvolution} may lead to a
symmetric hyperbolic system of equations for $m_{AB\mathcal{K}}$. A
special case of the above mentioned equation arises when
$M_{AB\mathcal{K}}$ has a symmetry relating the two components;

\item[(ii)] $m_{\mathcal{K}}$ cannot be reexpressed in terms of the
  $m_{AB\mathcal{K}}$ in which case the former is regarded as free
  data which has to be specified by means of a gauge choice. This may lead to
  a transport equation for $m_{AB\mathcal{K}}$. 

\end{itemize}

On the other hand, \eqref{ModelConstraint} is a constraint equation
for the components $m_{AB\mathcal{K}}$ which one expects to hold at
latter times if satisfied initially ---the so-called propagation of
the constraints. Note that, \emph{a priori}, there are no constraints
for $m_{\mathcal{K}}$.

\subsection{Hyperbolic reduction of a second  model equation}
\label{Section:ModelEquationAlt}

In the discussion of the propagation of the constraints a different
type of model equation will be considered. In what follows we will
briefly discuss its hyperbolic reduction.

\medskip
Let $M_{ij\mathcal{K}}=M_{[ij]\mathcal{K}}$ and
$N_{kij\mathcal{K}}=N_{[kij]\mathcal{K}}$ denote two tensorial
quantities, where again $\mathcal{K}$ stands for any set of spinor
indices. The model equation to be considered is given by
\begin{equation}
\hat{\nabla}_{[k} M_{ij]\mathcal{K}} = N_{kij\mathcal{K}}.
\label{AltModelEquation}
\end{equation}
This type of equation is motivated by the observation that if $\bm \omega$ is a  2-form, then its Lie derivative with respect to a vector field $\tau^\mu$ is given by
\[
\mathcal{L}_{\tau} {\bm \omega} = (i_\tau \mbox{d}+ \mbox{d} i_\tau) {\bm \omega},
\]
where $i_\tau {\bm \omega}$ denotes the contraction of the 2-form
${\bm \omega}$ with $\tau^\mu$. Now, if $i_\tau {\bm \omega}=0$ (as it
is the case with the the zero quantities associated with the extended
conformal field equations), one finds that
\[
\mathcal{L}_{\tau} \omega_{ij} = \tau^k \hat{\nabla}_{[k} \omega_{ij]}. 
\] 
In what follows, let $\epsilon_{ijkl}$ denote the components with
respect to the frame $e_k$ of the volume form of the metric
$g_{\mu\nu}$. Now,
\begin{eqnarray*}
&& \hat{\nabla}_{[k} M_{ij]\mathcal{K}}= \delta_{[k}{}^l \delta_i{}^m \delta_{j]}{}^n N_{lmn\mathcal{K}},\\
&& \phantom{\hat{\nabla}_{[k} M_{ij]\mathcal{K}}}= -\tfrac{1}{6} \epsilon_{kijp} \epsilon^{lmnp}N_{lmn\mathcal{K}}, \\
&& \phantom{\hat{\nabla}_{[k} M_{ij]\mathcal{K}}}= -\tfrac{1}{3} \epsilon_{kijp} \;{}^*\!N^{rp}{}_{r\mathcal{K}}.
\end{eqnarray*}
Because of the connection with the Lie derivative, it follows then that
\[
\tau^k\hat{\nabla}_{[k} M_{ij]\mathcal{K}} = -\tfrac{1}{3} \tau^k \epsilon_{kijp}\; {}^*\!N^{rp}{}_{r\mathcal{K}},
\]
implies an hyperbolic equation for the tensorial field
$M_{ij\mathcal{K}}$. The relevance of this equations to prove the
propagation of the constraints depends on whether its right hand side
can be casted as an homogeneous expression of other zero quantities ---see Section \ref{Section:PropagationOfConstraints}.

\subsection{The reduced geometric equations}

Following the discussion of the model equation \eqref{ModelEqn} in the
previous section one introduces the unprimed
spinorial fields $e^\es_{AB}$, $\hat{\Gamma}_{ABCD}$ and $\hat{P}_{ABCD}$
defined by 
\begin{eqnarray*}
&& e^\es_{AB} \equiv  \tau_{A}{}^{A'} e^\es_{BA'}, \\
&& f_{AB} \equiv \tau_{B}{}^{B'} f_{AA'}, \\
&& \hat{\Gamma}_{ABCD} \equiv \tau_B{}^{B'} \hat{\Gamma}_{AB'CD}, \\
&& \hat{P}_{ABCD}  \equiv  \tau_{B}{}^{A'} \tau_{D}{}^{C'} \hat{P}_{AA'CC'},
\\
&& \tilde{Y}_{ABCD}  \equiv  \tau_D{}^{C'}   \tilde{Y}_{ABCC'}
\end{eqnarray*} 
from which the original spacetime spinors $e^\es_{AA'}$ (the spinorial
version of $e^\es_k$), $f_{AA'}$, 
$\hat{\Gamma}_{AA'BC}$, $\hat{P}_{AA'BB'}$, $\tilde{Y}_{ABCC'}$  can be recovered using the
identity \eqref{TauTauEpsilon}. Following the discussion from the
previous section, the conformal equations \eqref{Xcfe1}-\eqref{Xcfe3}
can only yield evolution equations for the components
\[
e_{(AB)}, \quad f_{(AB)}, \quad \hat{\Gamma}_{(AB)CD}, \quad \hat{P}_{(AB)CD}.
\]
Due to the absence of further symmetries in the fields $e_{AB}$,
$f_{AB}$, $\hat{\Gamma}_{ABCD}$, $\hat{P}_{ABCD}$ the components
\[
e_Q{}^Q, \quad f_Q{}^Q, \quad \hat{\Gamma}_Q{}^Q{}_{CD}, \quad \hat{P}_Q{}^Q{}_{CD},
\]
are regarded as freely specifiable, and will be fixed by means of the
gauge conditions \eqref{GaussianGauge:Spinorial} so that in particular:
\[
f_Q{}^Q=0, \quad \hat{\Gamma}_Q{}^Q{}_{CD}=0, \quad \hat{P}_Q{}^Q{}_{CD} = \tilde{J}_{CD}.
\]
Consequently one obtains evolution equations of the form 
\begin{subequations}
\begin{eqnarray}
&&\partial_\tau e^\es_{(AB)} 
=  H^{[e]}_{AB}, \label{ReducedXcfe1}\\
&& \partial_\tau f_{(AB)} = H^{[f]}_{AB}, \label{ReducedXcfe2a}\\
&&\partial_\tau \hat{\Gamma}_{(AB)CD} 
= H ^{[\Gamma]}_{ABCD}, \label{ReducedXcfe2b}\\
&&\partial_\tau  \hat{P}_{(AB)CD}  
= \hat{\mathcal{D}}_{AB}  \tilde{J}_{CD}  +  \tilde{Y}_{ABCD} + H^{[P]}_{(AB)CD}. \label{ReducedXcfe3}
\end{eqnarray}
\end{subequations}
where $H^{[e]}_{AB}$ is a quadratic expression involving $e^\es_{(AB)}$
and $\hat{\Gamma}_{(AB)CD}$; $H^{[f]}_{AB}$ contains a quadratic
expression involving $f_{(AB)}$ and $\hat{\Gamma}_{(AB)CD}$ and a
linear term involving $\hat{P}_{(AB)CD}$; $H ^{[\Gamma]}_{ABCD}$
contains terms quadratic in $\hat{\Gamma}_{(AB)CD}$, a quadratic
expression involving $\Theta$ and $\phi_{ABCD}$ and a linear
expression in $\hat{P}_{(AB)CD}$; finally, $H^{[P]}_{(AB)CD}$ contains
quadratic terms in $\hat{\Gamma}_{(AB)CD}$ and $\hat{P}_{(AB)CD}$ and
in $d_{AB}$ and $\phi_{ABCD}$. Their explicit form will not be
important for our subsequent discussion.

\medskip
As it will be discussed in Section \ref{Section:PhysicalCotton}, the
spinor $\tilde{Y}_{ABCD}$ for the case of the Einstein-Maxwell system
contains derivatives of the Maxwell field. These terms enter in the
principal of equation \eqref{ReducedXcfe3}. This feature requires
 us to introduce new field equations ---essentially, the covariant
derivative of the  Maxwell spinor. The term $\hat{\mathcal{D}}_{AB}
\tilde{J}_{CD}$ in equation \eqref{ReducedXcfe3} will lead to similar
problems, for it will be seen that $\tilde{J}_{CD}$ is quadratic in
the Maxwell spinor. 

\medskip
For the evolution of the Jacobi field we split its space spinor into
irreducible components 
\[
\tau_B{}^{B'}\eta_{AB'} = \half \epsilon_{AB} \eta + \eta_{AB}, \quad\quad
 \mathrm{with} \quad \eta = \eta_A{}^A, \quad \eta_{AB}= \tau_{(B}{}^{B'}\eta_{A)B'}
\]
Then we get the evolution equations
\begin{subequations}
\begin{eqnarray}
&&\partial_\tau\eta = \half \eta \hat{\chi}_P{}^P{}_Q{}^Q + \eta^{CD} \hat{\chi}_{(CD)Q}{}^Q  , \\
&& \partial_\tau\eta_{AB} = \half \eta \hat{\chi}_P{}^P{}_{AB} + \eta^{CD} \hat{\chi}_{(CD)AB} .
\end{eqnarray}
\end{subequations}

\subsection{The reduced Bianchi equation}
In order to construct an evolution equation for the Weyl spinor  $\phi_{ABCD}$ we consider
the zero quantity
\begin{subequations}
\begin{eqnarray*}
&&\hat{\Lambda}_{ABCD}\equiv \tau_A{}^{A'} \hat{\Lambda}_{A'BCD}, \\
&& \phantom{\Lambda_{ABCD}} = \hat{\nabla}^Q{}_A \phi_{BCDQ}-f^Q{}_A
\phi_{BCDQ} -\Theta^{-1}\tilde{Y}_{BCDA}.
\end{eqnarray*}
\end{subequations}
Again, using the decomposition \eqref{SpaceSpinorNablaDecomposition}
one obtains
\[
\hat{\Lambda}_{ABCD} = - \tfrac{1}{2}\hat{\mathcal{P}} \phi_{ABCD}
+ \hat{\mathcal{D}}^Q{}_A \phi_{BCDQ} -f^Q{}_A
\phi_{BCDQ} -\Theta^{-1}\tilde{Y}_{BCDA}.
\]
Accordingly, 
\begin{equation}
-2 \hat{\Lambda}_{ABCD} = \hat{\mathcal{P}} \phi_{ABCD}
-2 \hat{\mathcal{D}}^Q{}_{(A} \phi_{BCD)Q} +2f^Q{}_{(A}
\phi_{BCD)Q} +2\Theta^{-1}\tilde{Y}_{(ABCD)}. \label{ReducedBianchi}
\end{equation}
renders the desired reduced equation.  In equation
\eqref{ReducedBianchi} we notice again the presence of the term
$\tilde{Y}_{(ABCD)}$ so that the same problem arises as for the
reduced equation \eqref{ReducedXcfe3}. We will thus treat this term in
the same way as outline for equation \eqref{ReducedXcfe3}, in order to
ensure the symmetric hyperbolicity of the system. We observe the
presence of the potentially singular term $\Theta^{-1}$. However, as
will be seen in the sequel, this term is cancelled out by a a similar
term appearing in the explicit form of $\tilde{Y}_{ABCD}$.

\medskip
Finally, it is noticed that the remaining content of the zero quantity $\Lambda_{ABCD}$
is contained in
\[
\hat{\Lambda}_P{}^P{}_{CD} = \hat{\mathcal{D}}^{PQ} \phi_{PQCD}- f^{PQ}\phi_{CDPQ}-\Theta^{-1}\tilde{Y}_{CD}{}^P{}_P, 
\]
corresponding to the constraints associated to the Bianchi identity
\eqref{Xcfe4}.

\section{The spinorial Maxwell equations}
\label{Section:MaxwellEquations}
Up to this point our discussion has been completely general and
irrespective of the trace-free matter models under consideration. In
order to proceed further, explicit information about the matter model
has to be provided ---in our case the Maxwell field.

\subsection{The Maxwell equations in the physical spacetime}
The physical spacetime Maxwell equations \eqref{EM-FE:3} are
equivalent to the spinorial equation
\begin{equation}
\label{PhysicalSpinorialMaxwell}
\tilde{\nabla}^{AA'} \tilde{\phi}_{AB}=0,
\end{equation}
where the antisymmetric Maxwell tensor $\tilde{F}_{\mu\nu}$ and the
totally symmetric spinor $\tilde{\phi}_{AB}$ are related to each other
by the correspondence
\[
\tilde{F}_{\mu\nu} \leftrightarrow \tilde{F}_{AA'BB'} \equiv \tilde{\phi}_{AB}\epsilon_{A'B'} + \bar{\tilde{\phi}}_{A'B'} \epsilon_{AB},
\]
with
\[
\tilde{\phi}_{AB} = \frac{1}{2} \tilde{F}_{AQ'B}{}^{Q'}. 
\]
The energy-momentum tensor \eqref{EM-FE:2} is given in spinorial terms by
\[
\tilde{T}_{AA'BB'} = \tilde{\phi}_{AB}\bar{\tilde{\phi}}_{A'B'}, 
\]
with
\[
\tilde{\nabla}^{AA'} \tilde{T}_{AA'BB'}=0.
\]

\subsection{The Maxwell equations in the unphysical spacetime}
If upon the conformal rescaling \eqref{rescaled:metric} one imposes the transformation rule
\begin{equation}
\label{rescaled:Maxwell}
\phi_{AB} = \Theta^{-1} \tilde{\phi}_{AB},
\end{equation}
then one obtains that
\begin{equation}
\label{MaxwellEquation}
\nabla^Q{}_{A'}\phi_{BQ}=0. \quad   
\end{equation}
In terms of the Weyl connection $\hat{\nabla}$ one has that
\begin{equation}
\label{WeylMaxwellEquation}
\hat{\nabla}^Q{}_{A'}\phi_{BQ} = f^Q{}_{A'}\phi_{BQ}.
\end{equation}
For later use we define the space spinor $\phi^\dagger_{AB}=
\tau_A{}^{A'} \tau_B{}^{B'} \bar{\phi}_{A'B'} $. For more details on
the Hermitian conjugation map for space spinors, see e.g. \cite{Fri95}.

\medskip
With regards to the stress-energy tensor one has that
\begin{equation}
T_{AA'BB'}\equiv \phi_{AB}\bar{\phi}_{A'B'}= \Theta^{-2}\tilde{T}_{AA'BB'}, \label{Maxwelltensortospinor}  
\end{equation}
with
\[
\nabla^{AA'}T_{AA'BB'}=0.
\]
This last property follows from the trace-freeness property of
$T_{AA'BB'}$ in four dimensions \cite{Fri91}. As a consequence of this
discussion, the following zero quantity is introduced:
\[
\hat{\omega}_{A'B} \equiv \hat{\nabla}^Q{}_{A'} \phi_{BQ} - f^Q{}_{A'}\phi_{BQ}.
\]

\medskip
In the sequel it will be seen that in order to obtain a symmetric
hyperbolic reduction of the conformal Einstein-Maxwell equations, it
is necessary to introduce the derivatives of the Maxwell field as a
variable. For this, one considers for a given gauge choice a spinorial field
$\hat{\psi}_{AA'BC}=\hat{\psi}_{AA'(BC)}$ and an associated zero
quantity $\hat{\omega}_{AA'BC}$.
These two quantities are related by
\[
\hat{\omega}_{AA'BC} \equiv \hat{\psi}_{AA'BC} - \hat{\nabla}_{AA'} \phi_{BC}.
\]
and under a connection change \eqref{ChangeOfConnection:2} the spinorial field $\hat{\psi}_{AA'BC} $ is adapted as
\begin{equation}
\label{transformation:psi}
\hat{\psi}_{AA'BC}=\psi_{AA'BC} - 2 \phi_{A(B} f_{C)A'}.
\end{equation}
The zero quantity $\hat{\omega}_{AA'BC}$ will be handled in the sequel
as a constraint. In order to obtain an equation for
$\hat{\psi}_{AA'BC}$ we adopt the strategy used in \cite{Fri91} and
make use of the Ricci identity ---cfr. equation
\eqref{GeneralRicciIdentity}--- for the Weyl connection $\hat{\nabla}$
applied to the spinor $\phi_{AB}$:
\[
\hat{\nabla}_{AA'} \hat{\nabla}_{BB'} \phi_{EF}-
\hat{\nabla}_{BB'} \hat{\nabla}_{AA'} \phi_{EF} = -2\phi_{P(E}
\hat{r}^P{}_{F)AA'BB'} -\hat{\Sigma}_{AA'}{}^{PP'}{}_{BB'} \hat{\nabla}_{PP'} \phi_{EF}.
\]
Replacing the derivatives of $\phi_{AB}$ by $\hat{\psi}_{AA'BC}$ and
assuming that the conformal equations \eqref{SpinorialXCFE} are satisfied one obtains the
required equation:
\begin{equation}
\hat{\nabla}_{AA'} \hat{\psi}_{BB'EF} -
\hat{\nabla}_{BB'} \hat{\psi}_{AA'EF} = -2\phi_{P(E}
\hat{R}^P{}_{F)AA'BB'}. \label{Equation:psi}
\end{equation}
To the latter we associate the following zero quantity:
\begin{equation}
\hat{\omega}_{AA'BB'EF} \equiv \hat{\nabla}_{AA'} \hat{\psi}_{BB'EF} -
\hat{\nabla}_{BB'} \hat{\psi}_{AA'EF} +2\phi_{P(E}
\hat{R}^P{}_{F)AA'BB'}.
\label{Definition:Omega}
\end{equation}

\medskip
\noindent
\textbf{Remark.} It can be verified from \eqref{rescaled:Maxwell} and \eqref{transformation:psi} that the zero quantities
$\hat{\omega}_{A'B}$, $\hat{\omega}_{A'ABC}$,
$\hat{\omega}_{AA'BB'CC'}$ transform homogeneously upon changes of
conformal gauge. This observation is of relevance for the propagation
of the constraints.  

\subsection{The reduced Maxwell equations}
The reduced equations implied by the Maxwell equations
\eqref{MaxwellEquation} is handled in a similar way to the Bianchi identity
\eqref{Xcfe4} ---one considers the unprimed zero quantity
\begin{eqnarray*}
&& \hat{\omega}_{AB} \equiv \tau_A{}^{A'} \hat{\omega}_{A'B}, \\
&& \phantom{\omega_{AB}} = \hat{\nabla}^Q{}_A \phi_{BQ}- f^Q{}_A \phi_{BQ}, \\
&&\phantom{\omega_{AB}} = -\tfrac{1}{2} \hat{\mathcal{P}} \phi_{AB} +
\hat{\mathcal{D}}^Q{}_A \phi_{BQ} - f^Q{}_A \phi_{BQ}.
\end{eqnarray*}
from where one obtains the propagation equation
\begin{equation}
-2 \hat{\omega}_{(AB)} = \hat{\mathcal{P}}\phi_{AB} -2
\hat{\mathcal{D}}^Q{}_{(A}\phi_{B)Q} +2 f^Q{}_{(A}\phi_{B)Q}=0,
\label{ReducedMaxwell}
\end{equation}
and the constraint
\[
\omega_Q{}^Q = \hat{\mathcal{D}}^{PQ}\phi_{PQ}-f^{PQ}\phi_{PQ}=0.
\]

\subsection{The reduced equations for the derivatives of the Maxwell spinor}
The treatment of the equation associated with the zero quantity
$\omega_{AA'BB'EF}$ follows the model discussed in Section
\ref{Section:ModelEquation}. In particular, one has to consider the
contracted zero quantity 
\[
\frac{1}{2} \omega_{(A|Q'|B)}{}^{Q'}{}_{EF} \equiv
\hat{\nabla}_{(A|Q'|} \hat{\psi}_{B)}{}^{Q'}{}_{EF} + 2 \phi_{P(E}\hat{R}^P{}_{F)}{}_{(A|Q'|B)}{}^{Q'}
\]
and its complex conjugate. The procedure described in Section
\ref{Section:ModelEquation} then leads to 
\begin{equation}
\hat{\mathcal{P}} \hat{\psi}_{(AB)CD} -\hat{\mathcal{D}}_{AB}
\hat{\psi}_P{}^P{}_{CD} = H^{[\psi]}_{ABCD}, \label{Evolution:psi}
\end{equation}
where $\hat{\psi}_{ABCD}$ is the space spinor version of
$\hat{\psi}_{AA'BC}$ given by
\[
\hat{\psi}_{ABCD} \equiv \tau_{B}{}^{A'} \hat{\psi}_{AA'CD}.
\]
The source term $H^{[\psi]}_{ABCD}$ contains quadratic terms involving
$\phi_{ABCD}$ and $\psi_{ABCD}$, $\phi_{AB}$ and $\hat{P}_{ABCD}$ and
$\phi_{AB}$ and $\phi_{ABCD}$. Some of parts of the quadratic
expression involving $\phi_{AB}$ and $\hat{P}_{ABCD}$ lead to terms
cubic in $\phi_{AB}$. As in the case of the reduced equations
\eqref{ReducedXcfe1}-\eqref{ReducedXcfe3}, the explicit form of the source
$H^{[\psi]}_{ABCD}$ will not be required.

\medskip
The reduction procedure described in the previous
lines does not provide an evolution equation
for the components $\hat{\psi}_P{}^P{}_{CD}$. To get around this, we
write
\begin{equation}
\hat{\psi}_{ABCD} =\nu_{ABCD} +\tfrac{1}{2} \epsilon_{AB} \nu_{CD} 
\label{Decomposition:psi}
\end{equation}
where
\[
\nu_{ABCD} \equiv \hat{\psi}_{(AB)CD}, \quad \nu_{CD} \equiv \hat{\psi}_P{}^P{}_{CD}.
\]
Let also
\[
\nu \equiv \hat{\psi}_{PQ}{}^{PQ}.
\]
It follows then that
\[
\hat{\psi}_{(ABCD)}= \nu_{(ABCD)} \quad
\hat{\psi}_{P(BC)}{}^P=\nu_{P(BC)}{}^P +  \tfrac{1}{2}
\nu_{BC}, \quad \hat{\psi}_{PQ}{}^{PQ}= \nu_{PQ}{}^{PQ}=\nu.
\]
Now, assuming that 
\[
\hat{\omega}_{AB}=0, \quad \hat{\omega}_{AA'BC}=0 
\]
so that $\hat{\psi}_{ABCD}=\hat{\nabla}_{AB}\phi_{CD}$ and the Maxwell
equations hold, one finds that
\begin{subequations}
\begin{eqnarray}
&& \nu_{ABCD} = {\hat{\cal{D}}}_{AB} \phi_{CD} \label{nu1}\\
&&\nu_{CD} = {\hat{\cal{P}}} \phi_{CD} =  - 2{{\hat{\cal{D}}}}_{Q(C} \phi_{D)}{}^Q - 2  f^Q{}_{(C}\phi_{D)Q} \nonumber\\
&&\phantom{\nu_{CD}}= -  2 \nu_{Q(CD)}{}^Q - 2  f^Q{}_{(C}\phi_{D)Q} \label{nu2}\\
&& \nu\tensor{AB}{AB}{} = {\hat{\cal{D}}}_{AB} \phi^{AB}. \label{nu3}
\end{eqnarray}
\end{subequations}
In particular, the relation \eqref{nu2} allows us to express the full
content of the field $\hat{\psi}_{ABCD}$ in terms of $\nu_{(ABCD)}$,
$\nu_{AB}$ and $\nu_{PQ}{}^{PQ}$ ---the term $\nu_{P(AB)}{}^P$ being
redundant. Substituting the decomposition \eqref{Decomposition:psi} into
equation \eqref{Evolution:psi} one obtains:
\begin{subequations}
\begin{eqnarray}
&& \hat{\mathcal{P}} \nu_{(ABCD)} -\hat{\mathcal{D}}_{(AB}\nu_{CD)} =
H^{[\psi]}_{(ABCD)}, \label{Reduced:psi1}\\
&& \hat{\mathcal{P}} \nu_{Q(BC)}{}^Q - \hat{\mathcal{D}}_{Q(B}
\nu_{C)}{}^Q = H^{[\psi]}_{Q(BC)}{}^Q. \label{quasiReduced:psi2} \\
&& \hat{\mathcal{P}} \nu -\hat{\mathcal{D}}_{PQ}\nu^{PQ} =
H^{[\psi]}_{PQ}{}^{PQ}. \label{Reduced:psi3}
\end{eqnarray}
\end{subequations}
Using equation \eqref{nu2} one finds that
\[
\hat{\mathcal{P}} \nu_{Q(BC)}{}^Q = -\tfrac{1}{2} \hat{\mathcal{P}} \nu_{BC} -
\hat{\mathcal{P}}f^Q{}_{(B} \phi_{C)Q} - f^Q{}_{(B} \hat{\mathcal{P}}\phi_{C)Q}.
\]
Using equation \eqref{ReducedXcfe2a} to replace $\hat{\mathcal{P}}f_{QB}$ by
$H^{[\psi]}_{AB}$ and equation \eqref{ReducedMaxwell} to express
$\hat{\mathcal{P}}\phi_{CQ}$ in terms of $\hat{\psi}_{ABCD}$ and a
quadratic expression in $f_{AB}$ and $\phi_{AB}$ it follows from
\eqref{quasiReduced:psi2} that
\begin{equation}
\hat{\mathcal{P}} \nu_{BC} + 2 \hat{\mathcal{D}}_{Q(B} \nu_{C)}{}^Q =
H^{\prime [\psi]}_{BC}, 
\label{Reduced:psi2}
\end{equation}
where $H^{\prime [\psi]}_{BC}$ contains the terms
appearing in $H^{[\psi]}_{BC}$ as well as a linear combination of the terms
appearing in $H^{[f]}_{AB}$ with terms quadratic in $f_{AB}$ and
$\hat{\psi}_{ABCD}$ and terms cubic in $f_{AB}$ and $\phi_{AB}$.

\medskip
Equations \eqref{Reduced:psi1}, \eqref{Reduced:psi3} together with
\eqref{Reduced:psi2} constitute a symmetric hyperbolic system of
equations for the components $\nu_{(ABCD)}$, $\nu_{AB}$ and $\nu$ of
the spinorial field $\hat{\psi}_{ABCD}$.

\subsection{The decomposition of the physical Cotton-York tensor}
\label{Section:PhysicalCotton}

In vacuum spacetimes the physical Cotton-York tensor $\tilde{Y}_{ijk}
$ vanishes. Thus, it does not appear in the Bianchi equations
\eqref{Bianchi}-\eqref{Bianchi2}.  In the case of trace-free matter
$\tilde{Y}_{ijk}$ is, in general, non-vanishing and carries
information about the physical fields in both equations \eqref{Bianchi} and
\eqref{Bianchi2}. The field $\tilde{Y}_{ijk} $ can be written in terms
of \emph{unphysical variables} as:
\begin{eqnarray*}
&&\tilde{Y}_{ijk} \equiv \tilde{\nabla}_{[i}\tilde{T}_{j]k} \\
&& \phantom{\tilde{Y}_{ijk}}=\Theta^2 \left( \hat{\nabla}_{[i} T_{j]k} + 3 b_{[i} T_{j]k} -2 f_{[i} T_{j]k} - g_{k[i}T_{j]m}g^{mn}b_n
\right)\nonumber.
\end{eqnarray*}
Substituting equations \eqref{MaxwellEquation} and \eqref{Maxwelltensortospinor}, 
recalling that $\eta_{AA'BB'}\equiv \epsilon_{AB}\epsilon_{A'B'}$ is the spinorial counterpart of $\eta_{ij}$ 
 and assuming that
$\hat{\psi}_{AA'BC} =\hat{\nabla}_{AA'} \phi_{BC}$ i.e. $\hat{\omega}_{AA'BC} = 0 $
one obtains 
\begin{eqnarray*}
&& \tilde{Y}_{AA'BB'CC'} =  \tfrac{1}{2}\Theta^2 \big(\hat{\psi}_{AA'BC} \bar{\phi}_{B'C'} + \bar{\hat{\psi}}_{AA'B'C'} \phi_{BC} -  \hat{\psi}_{BB'AC} \bar{\phi}_{A'C'} - \bar{\hat{\psi}}_{BB'A'C'} \phi_{AC}  \\
&&  \hspace{3.5cm} +(3 b_{AA'} - 2 f_{AA'})\phi_{BC}  \bar{\phi}_{B'C'} - (3 b_{BB'} - 2 f_{BB'})\phi_{AC}  \bar{\phi}_{A'C'} \\
&&  \hspace{3.5cm}- \epsilon_{CA}\epsilon_{C'A'} \phi_{BE}  \bar{\phi}_{B'E'}b^{EE'} + \epsilon_{CB}\epsilon_{C'B'} \phi_{AE}  \bar{\phi}_{A'E'}b^{EE'} \big).
\end{eqnarray*}
From the latter one readily finds that
\begin{eqnarray*}
&& \tilde{Y}_{ABCC'} \equiv \tfrac{1}{2}
\tilde{Y}_{AQ'B}{}^{Q'}{}_{CC'}, \\
&& \phantom{\tilde{Y}_{ABCC'}} = \tfrac{1}{2}\Theta^2 \big(
\hat{\psi}_{(A|Q'|B)C} \bar{\phi}^{Q'}{}_{C'} +
\phi_{C(A}\bar{\psi}_{B)Q'}{}^{Q'}{}_{C'} + 3\Theta^{-1} \phi_{C(A}
d_{B)Q'} \bar{\phi}^{Q'}{}_{C'} \\ 
&& \hspace{5.5cm} - 2 \phi_{C(A} f_{B)Q'} \bar{\phi}^{Q'}{}_{C'}
-\epsilon_{C(A}\phi_{B)E} \bar{\phi}_{C'E'} b^{EE'} \big).
\end{eqnarray*}
The corresponding unprimed version 
$\tilde{Y}_{ABCD}$ can be written entirely in terms of $\phi_{AB}$,
$\phi_{AB}^\dagger$, $\Theta$, $\dot{\Theta}$ and $d_{AB}$, by
recalling that 
\[
\bar{\phi}_{A'B'} = \tau^A{}_{A'} \tau^B{}_{B'} \phi^\dagger_{AB}. 
\] 
These explicit expressions will not be required in the subsequent
discussion. Important to note is that due to the presence of an
overall factor of $\Theta^2$ in $\tilde{Y}_{ABCC'}$, the term
$\Theta^{-1}\tilde{Y}_{ABCD}$ in equation \eqref{ReducedBianchi} is
formally regular at the points where $\Theta=0$. 

\subsection{Behaviour of the field variables and the zero quantities under gauge changes}

Before discussing the structural properties of the reduced conformal
Einstein-Maxwell equations in Section
\ref{Section:StructuralProperties} and the propagation of the
constraints in Section \ref{Section:PropagationOfConstraints} we would
like to briefly highlight the topic of gauge choice and gauge
invariance. The field variables and the zero quantities that have been
introduced in previous chapters have all been defined for a specific
choice of Weyl connection $\hat{\nabla}$, metric $g_{\mu\nu }$, frame
$\{ e_k \}$ and spinor dyad $\delta_A $ related by
\eqref{frame:metric}, \eqref{nabla_hat_of_g},
\eqref{ChangeOfConnection:2}.

Implied in their definitions are the transformation rules under gauge
change. These rules have either been explicitly given --
e.g. \eqref{SchoutenUnphysicalToWeyl2}, \eqref{rescaled:Maxwell},
\eqref{transformation:psi} -- or can be derived directly from these
rules and the extended conformal field equations
\eqref{XCFEFrame}. Therefore we refrain from listing them again.

However we would like to highlight that as a particular consequence of
these transformation rules, it follows that the various zero
quantities defined in earlier chapters are conformally covariant. Thus
if they vanish in one gauge they will also vanish in another. This
will be used in the discussion of the propagation of the constraint in
Section \ref{Section:PropagationOfConstraints}.

\section{Structural properties of the reduced conformal Einstein-Maxwell equations}
\label{Section:StructuralProperties}

We summarise the analysis of Sections
\ref{Section:HyperbolicReductions} and \ref{Section:MaxwellEquations}
in a form suitable for the applications that will be given in the sequel.

\medskip
We introduce the notation
\begin{eqnarray*}
&& {\bm \upsilon} \equiv \left(e^\es_{AB}, \hat{\Gamma}_{ABCD}, \hat{P}_{ABCD},\eta,\eta_{AB}\right),  \\
&& {\bm \phi}\equiv \left(\phi_{ABCD}\right), \\
&& {\bm \varphi}\equiv \left(\phi_{AB}\right), \\
&& {\bm \psi} \equiv \left(\psi_{ABCD}\right),
\end{eqnarray*}
where it is understood that ${\bm \upsilon}$, ${\bm \phi}$, ${\bm
\varphi}$ and ${\bm \psi}$ contain only the independent irreducible components of
the respective spinors. Let also
\[
\mathbf{u}\equiv ({\bm \upsilon},  {\bm \phi}, {\bm \varphi}, {\bm \psi})
\]
 In terms of
these quantities the propagation equations
\eqref{ReducedXcfe1}-\eqref{ReducedXcfe3} can be written as: 
\begin{equation}
\label{upsilon:propagation}
\partial_\tau {\bm\upsilon} = \mathbf{K}{\bm\upsilon} + \mathbf{Q}({\bm\upsilon},{\bm\upsilon}) +
\mathbf{R}({\bm\varphi},{\bm\psi}) + \mathbf{T}({\bm\phi},{\bm\psi},{\bm\upsilon})+ \mathbf{L}{\bm\phi},
\end{equation}
where $\mathbf{K}$ denotes a matrix with constant coefficients, $\mathbf{Q}({\bm\upsilon},{\bm\upsilon})$, $\mathbf{R}({\bm\varphi},{\bm\psi})$
bilinear vector value functions with constant coefficients and $\mathbf{T}({\bm\phi},{\bm\psi},{\bm\upsilon})$ a
trilinear vector valued function with constant coefficients. On the
other hand,  $\mathbf{L}$ is a linear matrix-valued function with
coefficients depending on the coordinates.  Equations  \eqref{ReducedBianchi}, \eqref{ReducedMaxwell} and \eqref{Reduced:psi1}-\eqref{Reduced:psi3}
can be written in the form
\begin{subequations}
\begin{eqnarray}
&&\hspace{-5mm} \left( \sqrt{2}\mathbf{E}_{5\times 5} + \mathbf{A}^{\underline{0}}_{5\times
    5}\right) \partial_\tau{\bm\phi} + \mathbf{A}^\er_{5\times
  5}\partial_\er {\bm \phi}
=\mathbf{B}({\bm\upsilon}){\bm\phi} + \mathbf{M}({\bm\psi},{\bm\varphi})+
\mathbf{N}({\bm \varphi},{\bm\varphi}), \label{phi:propagation} \\
&&\hspace{-5mm} \left( \sqrt{2}\mathbf{E}_{3\times 3} +\mathbf{A}^{\underline{0}}_{3\times 3}\right)\partial_\tau {\bm\varphi} + \mathbf{A}^\er_{3\times 3}\partial_\er {\bm\varphi}
=\mathbf{C}(\bm{\upsilon}){\bm\varphi},  \label{varphi:propagation}\\
&& \hspace{-5mm} \left( \sqrt{2}\mathbf{E}_{9\times 9} +\mathbf{A}^{\underline{0}}_{9\times 9}\right)\partial_\tau {\bm\psi} + \mathbf{A}^\er_{9\times 9}\partial_\er {\bm\psi}
=\mathbf{D}({\bm\upsilon}){\bm\psi}+
\mathbf{U}({\bm\upsilon},{\bm\varphi}) + \mathbf{V}({\bm
  \upsilon},{\bm \phi}) + \mathbf{W}({\bm \upsilon},{\bm \upsilon},{\bm
\phi}) ,  \label{psi:propagation}
\end{eqnarray}
\end{subequations}
where $E_{3\times 3}$, $E_{5\times 5}$, $E_{8\times 8}$
 denote,
respectively, the $3\times 3$, $5\times 5$ and $8\times 8$ identity
matrices, while $A^\es_{3\times 3}$, $A^\es_{5\times 5}$,
$A^\es_{9\times 9}$, $\es=0,\ldots,3$ 
are $3\times 3$, $5\times 5$
and $8\times 8$ Hermitian matrices depending on the coordinates. On
the other hand
$\mathbf{B}({\bm\upsilon})$, $\mathbf{C}(\bm{\upsilon})$, $\mathbf{D}({\bm\upsilon})$
denote constant matrix-valued linear function of
the entries of $\bm \upsilon$, while 
$\mathbf{M}({\bm\psi},{\bm\varphi})$, $\mathbf{N}({\bm
  \varphi},{\bm\varphi})$,  $\mathbf{U}({\bm\upsilon},{\bm\varphi})$, $\mathbf{V}({\bm
  \upsilon},{\bm \phi})$
denote bilinear functions with coordinate dependent
coefficients. Finally, $ \mathbf{W}({\bm \upsilon},{\bm \upsilon},{\bm
\phi})$ is a trilinear function. The Hermitian matrices
\[
\sqrt{2}\mathbf{E}_{3\times 3} + \mathbf{A}^{\underline{0}}_{3\times
    3}, \quad \sqrt{2}\mathbf{E}_{5\times 5} + \mathbf{A}^{\underline{0}}_{5\times
    5}, \quad \sqrt{2}\mathbf{E}_{9\times 9} + \mathbf{A}^{\underline{0}}_{9\times
    9} 
\]
imply real symmetric matrices if one decomposes the entries of $\bm
\phi$, $\bm \varphi$ and $\bm \psi$ into real and imaginary
parts. Hence, \eqref{upsilon:propagation} and \eqref{phi:propagation}
- \eqref{psi:propagation} give rise to a symmetric hyperbolic system
for $\mathbf{u} $. 

\section{Propagation of the constraints}
\label{Section:PropagationOfConstraints}

In this section we show that the conformal constraint equations
propagate by virtue of the conformal evolution equations, thus
implying a solution to the whole conformal field equations. More
precisely,

\begin{lemma}
\label{Lemma:PropagationConstraints}
Let $\mathcal{V}$ be an open subset of $\mathcal{S}$ and let
$\mathcal{U}$ be an open neighbourhood in $\mathcal{S}\times
[0,\infty)$. Assume that the unknowns $({\bm \upsilon},{\bm \phi},{\bm
\varphi},{\bm \psi})$ given on $\mathcal{U}$
represent a smooth solution of the reduced equations
\eqref{upsilon:propagation}, \eqref{phi:propagation},
\eqref{varphi:propagation} and \eqref{psi:propagation} for data on
$\mathcal{V}$ satisfying the Einstein-Maxwell conformal constraint equations. Let
$g_{\mu\nu}$ be the metric for which the frame obtained from the
unknowns $\bm \upsilon$ is orthonormal and let
$\mathcal{D}^+(\mathcal{V})\subset \mathcal{U}$ be the future domain
of dependence of $\mathcal{V}$ with respect to $g_{\mu\nu}$. Then the
conformal Einstein-Maxwell field equations
\begin{eqnarray*}
&& \hat{\Sigma}_{AA'}{}^{BB'}{}_{CC'}=0, \quad \hat{\Xi}_{ABCC'DD'}=0,
\quad \hat{\Delta}_{AA'BB'CC'}=0, \quad \hat{\Lambda}_{AA'BBCC'}=0, \\
&& \hat{\omega}_{A'A}=0, \quad \hat{\omega}_{AA'BC}=0, \quad \hat{\omega}_{AA'BB'CD}=0,
\end{eqnarray*}
 are satisfied on
$\mathcal{D}^+(\mathcal{V})$ by the fields  $({\bm \upsilon},{\bm \phi},{\bm
\varphi},{\bm \psi})$. Furthermore, the
metric 
\[
\tilde{g}_{\mu\nu} = \Theta^{-2} g_{\mu\nu}
\]
is a solution to the (physical) Einstein-Maxwell field equations on
\[
\{ p\in \mathcal{D}^+(\mathcal{V})\;|\; \Theta(p)\neq 0\}.
\]
\end{lemma}

The proof of this result follows a combination of the techniques
discussed in \cite{Fri91} and in \cite{Fri95}. We divide the proof in
several steps. In order to ease the presentation, in the subsequent discussion we
will use tensorial notation whenever possible. The discussion of the
propagation of the constraints follows the lines of the arguments
given in \cite{Fri91,Fri95}. This argument requires long
computations to obtain a complicated system of subsidiary equations
for the various zero quantities involved in the extended conformal
field equations. Since the argument is not particularly illuminating 
and for the sake of the presentation, we follow the
spirit of previous sections of the article and present a schematic
description of the procedure. In addition, we provide an alternative
argument for the propagation of the constraints based on the local
existence results of \cite{Fri91}. 

\begin{itemize}

\item[\textbf{(a)}] \textbf{Propagation of the constraints as a
consequence of the subsidiary equations.}

 In the subsequent discussion it will be assumed that the reduced
conformal field equations \eqref{ReducedXCFEa}-\eqref{ReducedXCFEb}
are satisfied. Furthermore, it will be assumed that the gauge
conditions \eqref{GaussianGauge:Tensorial} hold. 

\medskip
We define the following zero quantities associated to the conformal
gauge:
\begin{eqnarray*}
&& \delta_ k \equiv b_k -f_k -\Upsilon_k, \\
&& \gamma_{ij} \equiv \tfrac{1}{2} \tilde{T}_{ij} - \hat{P}_{ij} - \hat{\nabla}_i b_j
- \tfrac{1}{2} S_{ij}{}^{kl} b_k b_l, \\
&& \varsigma_{ij} \equiv \hat{P}_{[ij]} - \hat{\nabla}_{[i} f_{j]} -
\tfrac{1}{2}S_{[ij]}{}^{kl} f_k f_l, 
\end{eqnarray*}
where $\tilde{T}_{ij}$ is given by the matter model under
consideration. Under the assumption that
equations \eqref{ReducedXCFEa}-\eqref{ReducedXCFEb} and
\eqref{GaussianGauge:Tensorial} are satisfied, a computation along the
lines discussed in \cite{Fri95} shows that
\begin{eqnarray*}
&& \partial_\tau \delta_k = H^{[\delta]}_k, \\
&& \partial_\tau \gamma_{ij} = H^{[\gamma]}_{ij}, \\
&& \partial_\tau \varsigma_{ij} = H^{[\varsigma]}_{ij},
\end{eqnarray*}
where $H^{[\delta]}_k$ is an homogeneous expression in the
zero-quantities $\delta_k$, $\gamma_{ij}$, $\varsigma_{ij}$ and
$\hat{\Sigma}_i{}^{}_k$; $H^{[\gamma]}_{ij}$ is an homogeneous
expression in $\gamma_{ij}$; finally $H^{[\varsigma]}_{ij}$ is
homogeneous in $\hat{\Xi}{}^k{}_{lij}$. The explicit form of the
\emph{source terms} in the above equations and the evolution equations
for the other zero quantities to be discussed in the sequel will not be required in the
following discussion.

\medskip
The discussion of the propagation equations for the zero quantities
$\hat{\Sigma}_i{}^j{}_k$, $\hat{\Xi}^k{}_{lij}$ and
$\hat{\Delta}_{kij}$ follow the model of equation
\eqref{AltModelEquation} discussed in Section
\ref{Section:ModelEquationAlt}. In this case a lengthy computation
shows that
\begin{eqnarray*}
&& \mathcal{L}_\tau \hat{\Sigma}_i{}^j{}_k = H^{[\Sigma]}{}_i{}^j{}_k, \\
&& \mathcal{L}_\tau \hat{\Xi}^k{}_{lij} = H^{[\Xi]}{}^k{}_{lij}, \\
&& \mathcal{L}_\tau \hat{\Delta}_{kij} = H^{[\Delta]}{}_{kij},
\end{eqnarray*}
where $ H^{[\Sigma]}{}_i{}^j{}_k$ is a homogeneous expression in the
zero-quantities $\hat{\Sigma}_i{}^{}_k$ and $\hat{\Xi}^k{}_{lij}$;
$H^{[\Xi]}{}^k{}_{lij}$ is an homogeneous expression on the geometrical
equations zero-quantities $\hat{\Xi}^k{}_{lij}$, $\hat{\Delta}_{kij}$,
$\hat{\Sigma}_i{}^j{}_k$, $\hat{\Lambda}_{kij}$ and the gauge
zero-quantity $\delta_k$; finally $H^{[\Delta]}{}_{kij}$ is homogeneous
in the geometrical equations zero-quantities $\hat{\Delta}_{kij}$,
$\hat{\Sigma}_i{}^{}_k$, $\hat{\Lambda}_{kij}$, the gauge
zero-quantities $\gamma_{ij}$ and $\delta_{k}$, and the tensorial
counterpart of the spinorial matter
zero-quantities $\hat{\omega}_{A'A}$, $\hat{\omega}_{AA'BC}$,
$\hat{\omega}_{AA'BB'CD}$. 

\medskip
The construction of a propagation equation equation for the Bianchi
equation $\hat{\Lambda}_{kij}$ is slightly different. Following the
discussion in \cite{Fri95} one considers the quantity $\hat{\nabla}^k
\hat{\Lambda}_{kij}$. A lengthy manipulation using the definition of
$\hat{\Lambda}_{kij}$ and symmetries of the Weyl tensor shows that 
\[
\hat{\nabla}^k \hat{\Lambda}_{kij} = H^{[\Lambda]}_{ij},
\]
where $H^{[\Lambda]}_{ij}$ depends homogeneously on the geometrical zero-quantities
$\hat{\Xi}^k{}_{lij}$, $\hat{\Sigma}_i{}^j{}_k$, the gauge
zero-quantity $\varsigma_{ij}$, and the tensorial counterpart of the matter zero-quantities
$\hat{\omega}_{A'A}$, $\hat{\omega}_{AA'BC}$,
$\hat{\omega}_{AA'BB'CD}$.  Now, the spinorial counterpart of $\hat{\nabla}^k
\hat{\Lambda}_{kij}$ is given by $\hat{\nabla}^{PP'}
\hat{\Lambda}_{P'PBC}$. A space-spinor decomposition shows that the
components of $\hat{\Lambda}_{A'ABC}$ satisfy a symmetric hyperbolic
equation. In particular, for the Bianchi constraint one has that
\[
\hat{\mathcal{P}} \hat{\Lambda}_P{}^P{}_{AB}
-\hat{\mathcal{D}}^Q{}_{(A} \hat{\Lambda}_P{}^P{}_{B)Q} = H^{[\Lambda]}_{AB}, 
\]
where $H^{[\Lambda]}_{AB}$ has the same dependence on zero-quantities
as $H^{[\Lambda]}_{ij}$.

\medskip
Finally, for the constraints associated to the matter equations (the
Maxwell field) one has that an analogous procedure to the one
described in \cite{Fri91} renders also symmetric hyperbolic equations
for the components of $\hat{\omega}_{A'A}$, $\hat{\omega}_{AA'BC}$,
$\hat{\omega}_{AA'BB'CD}$ which are homogeneous in the matter
zero-quantities themselves and in the geometric zero-quantities. 

\medskip
Summarising: in the gauge
\eqref{GaussianGauge:Spinorial}  and as a consequence
of the reduced equations the geometrical zero-quantities
\[
\hat{\Sigma}_{AA'}{}^{BB'}{}_{CC'}, \quad \hat{\Xi}_{ABCC'DD'}, \quad \hat{\Delta}_{ABCC'}, \quad \hat{\Lambda}_{A'ABC}, 
\]
together with the Maxwell zero-quantities
\[
\hat{\omega}_{A'A}, \quad \hat{\omega}_{AA'BC}, \quad \hat{\omega}_{AA'BB'CD}
\]
and the gauge zero-quantities 
\[
\delta_{AA'}, \quad \gamma_{AA'BB'}, \quad \varsigma_{AA'BB'}
\]
form a symmetric hyperbolic system which is homogeneous in the zero
quantities themselves. Accordingly, if
the zero-quantities vanish on $\mathcal{V}\subset\mathcal{S}$, then
the zero-quantities vanish on $\mathcal{D}^+(\mathcal{V})$. Hence one
has a solution to the conformal Einstein-Maxwell field equations of
$\mathcal{D}^+(\mathcal{V})$.

\item[\textbf{(a')}] \textbf{An alternative argument for the propagation of the constraints.}

An argument involving less computations to prove the propagation of the constraints can be
obtained by directly exploiting the local existence results of
\cite{Fri91}. In what follows we consider the extended conformal field
equations \eqref{Xcfe1}-\eqref{Xcfe4} in an arbitrary gauge. We notice
that if one sets $f_{AA'}=0$ in these equations, then from equation
\eqref{defining di} it follows that $d_{AA'} = \nabla_{AA'} \Theta$,
and hence, the arbitrary Weyl connection $\hat{\nabla}$ reduces to the
associated Levi-Civita connection $\nabla$. In order to obtain the
full correspondence with the conformal equations of \cite{Fri91} one
has to prescribe equations for the conformal factor $\Theta$
and the 1-form $d_{AA'}$.  The relevant
equations are given by
\begin{subequations}
\begin{eqnarray}
&&  \nabla_{AA'} d_{BB'} =\nabla_{AA'} \nabla_{BB'}\Theta=-\Theta P_{AA'BB'} + s\, \epsilon_{AB} \epsilon_{A'B'} + \tfrac{1}{2} \Theta \, \tilde{T}_{AA'BB'}, \label{LC:ConditionA}\\
&& \nabla_{AA'} s = -P_{AA'BB'} \nabla^{BB'} \Theta + \tfrac{1}{2} \tilde{T}_{AA'BB'}\nabla^{BB'} \Theta, \label{LC:ConditionB} \\
&& 6 \Theta\, s - d_{PP'} d^{PP'}= \tilde{\lambda}. \label{LC:ConditionC}
\end{eqnarray}
\end{subequations}
with
\[
s \equiv \tfrac{1}{4} \nabla_{PP'} \nabla^{PP'}\Theta + \tfrac{1}{24} R\, \Theta. 
\]
Equations \eqref{LC:ConditionA}-\eqref{LC:ConditionC} arise from the
transformation rule for the Schouten tensor under conformal rescalings
and from the definition of $s$.  It should be noted that these
equations are not required in the
particular type of hyperbolic reduction considered in this article as
both $\Theta$ and $d_{AA'}$ are fixed by the gauge of Section
\ref{Section:DefinitionGauge} ---the generalised conformal Gaussian systems.

\medskip
From the theory in \cite{Fri91} one has that given initial data
satisfying the conformal constraint equations on
$\mathcal{V}\subset\mathcal{S}$, there exists $\mathcal{W}\subset
\mathcal{D}^+(\mathcal{V})$ 
in which the ``standard'' conformal field
equations are satisfied. In particular, this implies the existence of
a physical spacetime with metric $\tilde{g}_{\mu\nu}$. The solution
constructed by the procedure of \cite{Fri91} is unique up to conformal
rescalings, coordinate transformations and a choice of frame. For the
subsequent discussion we denote the conformal factor, metric and
associated Levi-Civita
connection thus obtained by $\check{\Theta}$, $\check{g}_{\mu\nu}$ and
$\check{\nabla}$. The metrics $\check{g}_{\mu\nu}$ and
$\tilde{g}_{\mu\nu}$ are related via 
\[
\check{g}_{\mu\nu}=\check{\Theta}^2 \tilde{g}_{\mu\nu}.
\]

\medskip
Now, consider on $\mathcal{V}\subset \mathcal{S}$ initial data for the 
$\check{\nabla}$-version of the equations \eqref{Gcg1}-\eqref{Gcg2} to construct a
timelike congruence of conformal curves. It follows from standard theorems on
the existence of ordinary differential equations that given a set of
initial data for these curves, one can always find a smaller subset $\mathcal{W}'\subset
\mathcal{W}$ such that the congruence is free of conjugate
points. Thus, through every point $p \in  \mathcal{U}'$ we have a
unique conformal curve starting on $\mathcal{V}$. In particular, one
can choose initial data for the congruence so that one obtains a
generalised conformal Gaussian system like the one described in
Section \ref{Section:ConformalCurves}.  The solution to the
$\check{\nabla}$-version of the conformal curve equations gives a
1-form $\check{b}_\mu$  on $\mathcal{W}'$. Furthermore, recall that
given initial data on $\mathcal{V}$ one can construct a preferred
conformal factor $\check{\mho}$ by solving the appropriate version of
equation \eqref{Theta evolution}:
\[
\dot{\check{\mho}} = \check{\mho} \langle \check{b}, v\rangle.
\]
In particular, one may choose as
initial data $\check{\mho}_*=1$ on $\mathcal{V}$. This choice connects the metric
$\check{g}_{\mu\nu}$ with the unique metric $g_{\mu\nu}$ for which the
tangent vectors $v^\mu$ to the congruence of conformal curves are taken to be orthogonal to
$\mathcal{S}$ and satisfy
\[
g(v, v)=1.
\]
One can use $\check{b}$ and $\check{\mho}$ to construct a Weyl
connection $\hat{\nabla} = \check{\nabla} + S(\check{b}) $ and the
associated Levi-Civita connection $\nabla =\check{\nabla} + S(\check{\mho}^{-1} \mbox{d}\check{\mho})$.

\medskip
By construction, the standard conformal field equations, and hence the
extended conformal field equations, are satisfied in the
gauge of the connection $\check{\nabla}$. As observed in Sections
\ref{Section:WeylConnections},\ref{Section:XCFE} and \ref{Section:MaxwellEquations} the zero quantities used
in the formulation of the extended conformal field equations are conformally
covariant. Thus, if the extended conformal field equations are
satisfied in the gauge $\check{\nabla}$, then they are also satisfied
in that given by $\hat{\nabla} $. As a consequence, the reduced
conformal field equations associated to the constructed congruence
hold. Since the solution to the reduced conformal field equations is
unique, it follows that using the above initial data one must obtain
the same solution for \eqref{upsilon:propagation}, \eqref{phi:propagation},
\eqref{varphi:propagation} and \eqref{psi:propagation} as the one
constructed above. In particular, one can finally conclude that the
constraint equations must be satisfied throughout $\mathcal{W}'$.

\item[\textbf{(b)}] \textbf{A solution to the conformal Einstein-Maxwell field
  equations implies a solution to the physical Einstein-Maxwell field equations. }

Assume now that on $\mathcal{U}$ one has a solution to the extended
Einstein-Maxwell conformal field
equations ---that is,
\begin{eqnarray*}
&& \hat{\Sigma}_{AA'}{}^{BB'}{}_{CC'} =0, \quad \hat{\Xi}_{ABCC'DD'} =0, \quad
\hat{\Delta}_{AA'BB'CC'}=0,\quad
\hat{\Lambda}_{A'BCD}=0, \\
&& \hat{\omega}_{A'A}=0, \quad \hat{\omega}_{AA'BB'CD}=0.
\end{eqnarray*}
Assume also that the additional zero quantities satisfy
\[
\hat{\omega}_{AA'BC}=0, \quad \delta_{AA'}=0, \quad  \gamma_{AA'BB'}=0, \quad \varsigma_{AA'BB'}=0.
\]
The solution to the reduced conformal field equations provides, in
particular, fields
\[ 
e_{AA'}, \quad f_{AA'} \quad \hat{\Gamma}_{AA'}{}^{BB'}{}_{CC'} \quad 
\mbox{ on } \mathcal{U},
\]
where in this discussion the 1-form $f_{AA'}$ is defined via
\[
 f_{AA'} \equiv \hat{\Gamma}_{AA'}{}^Q{}_Q .
\]
as $\hat{\Sigma}_{AA'}{}^{BB'}{}_{CC'}=0$. The connection coefficients
$\hat{\Gamma}_{AA'}{}^{BB'}{}_{CC'}$ give rise to a torsion-free
connection $\hat{\nabla}$. Motivated by the relation
\eqref{frame:metric} one can use the frame $e_{AA'}$ and the frame
metric $\eta_{AA'BB'} \equiv \epsilon_{AB}\epsilon_{A'B'}$ to
construct a metric $g_{\mu\nu}$. By construction
$e_{AA'}(\epsilon_{CD})=0$ so that
\[
\hat{\nabla}_{AA'}\epsilon_{CD} = -\hat{\Gamma}_{AA'}{}^Q{}_Q\epsilon_{CD}=-f_{AA'} \epsilon_{CD},
\]
which is the spinorial counterpart of 
\[
\hat{\nabla}_\mu g_{\nu\lambda} = -2 f_\mu g_{\nu\lambda}.
\]
Thus, $\hat{\nabla}$ is a Weyl connection for the metric
$g_{\mu\nu}$. Motivated by \eqref{WeylToUnphysical}, one defines the
connection $\nabla$ with connection coefficients
\[
\Gamma_{AA'}{}^{CC'}{}_{BB'} \equiv \hat{\Gamma}_{AA'}{}^{CC'}{}_{BB'} - S_{AA'BB'}{}^{CC'PP'} f_{PP'}.
\]
then 
\[
\nabla_{AA'}\epsilon_{CD} =0,
\]
so that $\nabla_\mu g_{\nu\lambda} =0 $ ---that is, $\nabla$ is a
metric connection. Using the invariance of the torsion under change of
connection ---cfr. equation \eqref{ConformalInvariance:Torsion}--- it
follows that $\nabla$ is torsion free. Thus, because of uniqueness, 
$\nabla$ must be the Levi-Civita connection of $g_{\mu\nu}$.

\medskip
Now, from
\[
\hat{\Xi}_{ABCC'DD'}\equiv \hat{r}_{ABCC'DD'} - \hat{R}_{ABCC'DD'}=0,
\]
 the fields $\hat{P}_{AA'BB'}$ and $\phi_{ABCD}$ on $\mathcal{U}$ obtained
as a solution of the reduced conformal field equations can be
identified, respectively, with the Schouten and Weyl spinors of the
Weyl connection $\hat{\nabla}$ ---recall that the
decomposition in terms of irreducible components is unique. 
Due to conformal invariance, the Weyl tensor of the Weyl connection $\hat{\nabla}$ is also the Weyl tensor of
the Levi-Civita connection $\nabla$. 

\medskip
Motivated by the rescaling \eqref{rescaled:metric} we  use the
transformation rule \eqref{ChangeOfConnection:1} to define a
\emph{physical} connection $\tilde{\nabla}$. From $\delta_{AA'}=0$ one
has that
\[
b_{AA'}= \Upsilon_{AA'} + f_{AA'},
\]
and accordingly $\tilde{\nabla} $ is the Levi-Civita connection of the
metric $\tilde{g}_{\mu} \equiv \Theta^{-2}g_{\mu\nu} $.  Using the
transformation rule \eqref{SchoutenPhysicalToWeyl2} one finds that
the physical Schouten spinor is given by 
\[
\tilde{P}_{AA'BB'} \equiv \hat{P}_{AA'BB'} + \hat{\nabla}_{AA'}\left( f_{BB'} + \Upsilon_{BB'} \right) + \tfrac{1}{2} S^{PP'QQ'}{}_{AA'BB'} \left( f_{PP'} + \Upsilon_{PP'} \right)\left( f_{QQ'} + \Upsilon_{QQ'} \right). 
\] 
Note that since $\varsigma_{AA'BB'}=0$, one has that 
\[
\tilde{P}_{AA'BB'}-\tilde{P}_{BB'AA'}=0, \quad \tilde{P}_{[ij]}=0.
\]
Furthermore, from $\gamma_{AA'BB'}=0$ one finds that
\begin{equation}
\tilde{P}_{AA'BB'} = \tfrac{1}{2} \Tilde{T}_{AA'BB'}, \quad  \tilde{P}_{ij} = \tfrac{1}{2} \Tilde{T}_{ij}.
\label{AlmostEMFE}
\end{equation}
From the field equations $\hat{\omega}_{A'A}=0$, $\hat{\omega}_{AA'BB'CC'}=0$ and the
constraint $\hat{\omega}_{AA'BC}=0$, one has that $\tilde{\phi}_{AB}$
satisfies the physical Maxwell equations. Thus, $\tilde{T}_{AA'BB'}$
defined by
\[
\tilde{T}_{AA'BB'} = \tilde{\phi}_{AB} \tilde{\phi}_{A'B'},
\]
is the energy momentum tensor of the Maxwell field and the equations
given in \eqref{AlmostEMFE}
are equivalent to the Einstein-Maxwell field equations.

\end{itemize}


\section{A first application: stability of Einstein-Maxwell de
  Sitter-like spacetimes}
\label{Section:deSitterSpacetimes}

The use of a gauge based on the conformal curves described in section
\ref{Section:ConformalCurves} allows to directly transcribe the analysis of the conformal
boundary for vacuum de Sitter-like spacetimes to the case of 
Einstein-Maxwell de Sitter-like spacetimes. 

\medskip
For the Sitter-like spacetimes one can formulate two slightly different
Cauchy initial value problems: one where initial data is prescribed on
a standard Cauchy hypersurface, and a second one where the data is
prescribed on one portion of the conformal boundary ---say, past null
infinity.  The de Sitter-like spacetimes
that will be considered have Cauchy slices with the topology of
$\Sphere^3$. The construction of suitable  coordinate systems and a
frame vectors this type of configurations has been discussed in detail
in \cite{LueVal09a,LueVal09b}. 

\subsection{Structure of the conformal boundary}
Following the general ideas of \cite{LueVal09a}, here we present a
brief discussion of the structure of the conformal boundary of de
Sitter-like Einstein-Maxwell spacetimes. 

\subsubsection{Standard Cauchy problem}
If the initial hypersurface $\mathcal{S}$ is a standard Cauchy
hypersurface one has that 
\[
\Theta = \Theta_*\left( 1+\tau \langle b_*, v_*\rangle
  +\frac{1}{2}\tau^2 \left(\tilde{\lambda} \Theta_*^{-2} +\frac{1}{2} g^\sharp(b_*,b*) \right) \right),
\]
for some $\Theta_*\neq 0$. The conformal factor vanishes at
\[
\tau_\pm = \frac{-2\Theta_*\langle d, v \rangle_* \pm 2\Theta_*
  \sqrt{|2\tilde{\lambda}+ g^\sharp(d,d)_*|}}{2\tilde{\lambda}+g^\sharp(d,d)_*}.
\]
One has then that
\begin{equation}
\mathscr{I}^\pm =\{ \tau_\pm \} \times \mathcal{S}.
\label{Scri1}
\end{equation}
Furthermore, $\nabla_k \Theta \nabla^k \Theta = -2\tilde{\lambda}$, so
that both components of null infinity are space-like. 

\subsubsection{Cauchy problem on past null infinity}
In the case of an initial value problem prescribed on null infinity,
one has that $\Theta_*=0$ so that 
\[
\Theta = \langle d, v \rangle_* \tau + \frac{1}{2} \ddot{\Theta}_*\tau^2
\]
Combining \eqref{LC:ConditionC} and Lemma \ref{Lemma:d} one finds that $g^\sharp(d,d)_* =-2\tilde{\lambda}$ and, if
one sets $d_*=(\nabla \Theta)_*$, that
\[
d_k(\tau) =\left( \sqrt{-2\tilde{\lambda}} + \ddot{\Theta}_*\tau,0,0,0 \right).
\]
The conformal factor vanishes at
\begin{equation}
\mathscr{I}^- =\{\tau=0\} \times \mathcal{S}, \quad \mathscr{I}^+ =
\left\{ \tau = -2 \dot{\Theta}_*/\ddot{\Theta}_*\right\}\times \mathcal{S}.
\label{Scri2}
\end{equation}
Note that the location of $\mathscr{I}^+$ is determined by the free
data $\ddot{\Theta}_*$.

\subsection{Stability of Einstein-Maxwell de Sitter-like spacetimes}

Combining the \emph{a priori} knowledge on the structure of the conformal
boundary discussed in the previous sections with the structural properties
of the reduced equations \eqref{upsilon:propagation},
\eqref{phi:propagation}, \eqref{varphi:propagation},
\eqref{psi:propagation} discussed in Section
\ref{Section:StructuralProperties}, Lemma
\ref{Lemma:PropagationConstraints} on the propagation of the
constraints, and Kato's existence and stability theorems for symmetric
hyperbolic systems \cite{Kat70,Kat73,Kat75} one obtains the following
existence and stability result for de Sitter-like Einstein-Maxwell
spacetimes. The proof is identical to that in
\cite{LueVal09a,LueVal09b} and it is omitted. Let in what follows
$\mathring{\mathbf{u}}$ denote the solution to the reduced equations
\eqref{upsilon:propagation}, \eqref{phi:propagation},
\eqref{varphi:propagation}, \eqref{psi:propagation} corresponding to
the (vacuum) de Sitter spacetime.

\begin{theorem}
Let $\mathbf{u}_0=\mathring{\mathbf{u}}_0 + \breve{\mathbf{u}}_0$ be Einstein-Maxwell Cauchy
(standard or at past null infinity) data for a de Sitter-like
spacetime. There exists $\varepsilon>0$ such that if $\breve{u}_0$ is
sufficiently small, then there exists on $[\tau_-,\tau_+]\times
\mathcal{S}$ a unique smooth solution $\mathbf{u}=\mathring{\mathbf{u}}+\breve{\mathbf{u}}$ to
the conformal propagation equations  \eqref{upsilon:propagation}, \eqref{phi:propagation},
\eqref{varphi:propagation}, \eqref{psi:propagation}  such that the associated
congruence of conformal curves contains no conjugate points in
$[\tau_-,\tau_+]$. The field $\mathbf{u}$ implies a smooth solution
to the Einstein-Maxwell field equations with positive cosmological
constant for which the sets $\mathscr{I}^\pm$ defined by \eqref{Scri1} ---in the
standard Cauchy problem--- or by \eqref{Scri2} ---in the Cauchy problem with data
at null infinity--- represent past and future null infinity.
\end{theorem}

\noindent
\textbf{Remark.} Note that this stability result for Einstein-Maxwell
spacetimes is given with respect to a vacuum reference spacetime.

\section{A second application: stability of Einstein-Maxwell radiative
  spacetimes}
\label{Section:RadiativeSpacetimes}
As a second example of our approach, we obtain a generalisation of
the stability results for purely radiative spacetimes discussed in
\cite{LueVal09b}. In contrast to the stability proof for de
Sitter-like Einstein-Maxwell spacetimes, in this case the reference
solution has a non-vanishing electromagnetic field. For the sake of
conciseness most of the technical details are omitted and we only
remark on those aspects of the analysis that differ from the treatment
for vacuum spacetimes given in \cite{LueVal09b}. 

\subsection{Einstein-Maxwell initial data sets with vanishing mass}

In what follows, a static solution to the Einstein Maxwell
solutions (an electrostatic solution) will be understood to be a triple
$(\tilde{h}_{\alpha\beta}, \tilde{\Phi},\tilde{\Psi})$, solving the
\emph{electrostatic field equations}. The (negative definite)
Riemannian 3-metric $\tilde{h}_{\alpha\beta}$ is the metric of the
quotient manifold, and $\Phi$, $\Psi$ denote, respectively, the
gravitational and electric potentials. Any static, asymptotically
flat solution to the Einstein-Maxwell equations admits an analytic compactification of a neighbourhood of
spatial infinity $i$ ---see \cite{Sim89}. The triple
$(\tilde{h}_{\alpha\beta}, \tilde{\Phi},\tilde{\Psi})$ can be suitably
rescaled to render another triple $(h_{\alpha\beta},\Phi,\Psi)$
which is  analytic in a neighbourhood $\mathcal{B}_a(i)$ and solves 
the \emph{conformal electrostatic field equations}. Any such triple
gives rise to a solution $(\bar{h}_{\alpha\beta},\bar{\Omega},\bar{E}_\alpha)$
of the (conformally rescaled) time symmetric Einstein Maxwell
constraints
\begin{eqnarray*}
&& \bar{r} = 2\bar{\Omega}^2 \bar{E}_\alpha \bar{E}^\alpha + 8 \bar{\Omega}^{1/2}
\bar{\D}_\alpha \bar{D}^\alpha(\Omega^{-1/2}), \\
&& \bar{D}_\alpha \bar{E}^\alpha =0.
\end{eqnarray*}
 with
vanishing mass and charge ---here $\bar{D}$ and $\bar{r}$ denote,
respectively, the Levi-Civita connection and Ricci scalar of the
metric $\bar{h}_{\alpha\beta}$; the tensor $\bar{E}_\alpha$ is the
electric field. From
$(\bar{h}_{\alpha\beta},\bar{\Omega},\bar{E}_\alpha)$ one can
construct initial data for the extended conformal field equations. In
particular, data for the Schouten and Weyl tensors are given,
respectively by the expressions
\begin{eqnarray*}
&& \bar{P}_{\alpha\beta} = -\bar{\Omega}^{-1}\mathcal{C}\left(\bar{D}_\alpha \bar{D}_\beta
  \bar{\Omega} \right)
-\bar{\Omega}^2 \mathcal{C}\left(  \bar{E}_\alpha \bar{E}_\beta\right) +\tfrac{1}{12} \left(  \bar{r} -2\bar{\Omega}^2 \bar{E}_\alpha \bar{E}^\alpha - 8 \bar{\Omega}^{1/2}
\bar{\D}_\alpha \bar{D}^\alpha(\bar\Omega^{-1/2})\right)h_{\alpha\beta}, \\
&& \bar{d}_{\alpha\beta} = \bar{\Omega}^{-2}\mathcal{C}(\bar{D}_\alpha \bar{D}_\beta\bar{\Omega}) + \bar{\Omega}^{-1}\mathcal{C}(\bar{r}_{\alpha\beta}) + \bar{\Omega}
\mathcal{C}(\bar{E}_\alpha \bar{E}_\beta),
\end{eqnarray*}
which can be shown to be analytic in $\mathcal{B}_a(i)$. In these last expressions,
$\mathcal{C}$ denotes the trace-free part of the tensor in
parenthesis. 

\subsection{Construction of a reference radiative Einstein-Maxwell spacetime}

Let $(\bar{h}_{\alpha\beta},\bar{\Omega},\bar{E}_\alpha)$ on $\bar{\mathcal{S}}$ be
one of the solutions to the time symmetric conformal 
constraint discussed in the
previous subsection. For the present purposes it will be convenient to
consider a conformal factor $\bar{\Omega}$ which is negative ---this obtained by making
the obvious sign changes in the relevant equations. By construction
$\bar{\Omega}$ satisfies the following \emph{asymptotic flatness conditions}:
\begin{equation}
\label{asymptotic_behaviour}
\bar{\Omega} < 0 \,\,\mathrm{on}\, \bar{\mathcal{S}}\setminus i , \quad
 \bar{\Omega}(i) = 0  , \quad 
D_{{\cal{A}}}\bar{\Omega}(i) = 0  , \quad 
D_{{\cal{A}}}D_{{\cal{B}}}\bar{\Omega}(i) = 2 h_{{\cal{A}}{\cal{B}}}(i). 
\end{equation}
We work in a suitably small neighbourhood,
$\mathcal{B}_a(i)\subset\bar{\mathcal{S}}$ such that all the
statements made in the sequel make sense. We use the coordinates
$x^{\cal{A}}, \, {\cal{A}}=1,2,3$ centred at $i$ and consider the
following initial data for a congruence of conformal curves:
\begin{equation}
\label{cg_initial_data}
\bar{\tau}_* = 0 , \quad \dot{x}^\mu = \bar{n}^\mu, \quad \bar{\Theta}_* = \bar{\Omega} , \quad \dot{\bar{\Theta}}_* =0, \quad \bar{d}_* \equiv \bar{\Theta}_* \bar{b}_* =(\mathrm{d}\bar{\Theta})_*.
\end{equation}
The coordinates $x^{\cal{A}}$ are extended off $\bar{\mathcal{S}}$ by
dragging along the congruence of conformal curves to obtain
generalised conformal Gaussian coordinates. It can be readily verified
that $\ddot{\bar{\Theta}}>0$ on $\bar{\mathcal{S}}$. It follows that along each conformal curve the conformal
factor $\bar{\Theta}$ is given by
\begin{equation}
\label{special:Theta}
\bar{\Theta}(\bar{\tau}) = \bar{\Omega} + \frac{1}{2}
\ddot{\bar{\Theta}}_* \bar{\tau}^2 = \bar{\Omega}\left(1 -
  \frac{\bar{\tau}^2}{\bar{\omega}^2} \right), \quad \bar{\omega} \equiv  \sqrt{\frac{2\bar{\Omega}}{\ddot{\bar{\Theta}}_*}}, \quad \bar{\omega} (i)=0.
\end{equation}
Define now the conformal boundary, $\bar{\scri}$, in a natural way as
the locus of points in the development of the data on
$\bar{\mathcal{S}}$ for which $\bar{\Theta}=0$ and
$\mbox{d}\bar{\Theta}\neq 0.$ It is easy to see that a conformal
curve with data given by (\ref{cg_initial_data}) passes through
$\bar{\mathscr{I}}$ whenever $\bar{\tau} =\pm \bar{\omega}$. 

\bigskip
Having located the conformal boundary for the evolution of data on
$\mathcal{B}_a(i)$ for the Einstein-Maxwell system, one can discuss
now the existence of solutions to the propagation system given by 
\eqref{upsilon:propagation}, \eqref{phi:propagation},
\eqref{varphi:propagation}, \eqref{psi:propagation}.  For this we extend the data on $\mathcal{B}_a(i)$
to data on the whole of $\bar{\mathcal{S}}\simeq \Sphere^3$ in the way
discussed in \cite{Fri86b,LueVal09a}. Using the same methods as in
\cite{LueVal09b} and Lemma \ref{Lemma:PropagationConstraints} one obtains the following local result:

\begin{theorem}
\label{Theorem:Local}
Given radiative data for the conformal Einstein-Maxwell equations,
there exist a $\bar{T}_0>0$ and on $\mathring{\mathcal{M}}\equiv
[-\bar{T}_0,\bar{T}_0]\times \bar{\mathcal{S}}$ a unique smooth
solution $\mathbf{u}$ to the propagation equations \eqref{upsilon:propagation}, \eqref{phi:propagation},
\eqref{varphi:propagation}, \eqref{psi:propagation}.  The solution $\mathbf{u}$
implies a solution to the conformal Einstein-Maxwell equations on
\[
\mathcal{M}  \equiv  \mathring{\mathcal{M}}  \cap I^-(i) \subset
\mathcal{D}(\mathcal{B}_a(i)).
\]
The spacetime $(\mathcal{M}\setminus \bar{\mathscr{I}},
\mathring{g}_{\mu\nu} ) $ implied by the solution to the conformal
Einstein-Maxwell field equations is conformally related to an
Einstein-Maxwell spacetime spacetime, $(\mathcal{M}\setminus
\bar{\mathscr{I}}, \bar{\Theta}^{-2}\mathring{g}_{\mu\nu})$, with
vanishing cosmological constant. The spacetime $(\mathcal{M}\setminus
\bar{\mathscr{I}}, \Theta^{-2} \mathring{g}_{\mu\nu} ) $ is a
radiative spacetime for which the set $\bar{\mathscr{I}}^+$
corresponds to its future null infinity, while the point 
$i^+=(0,i)\in \{0\} \times \bar{\mathcal{S}}$ is its future timelike
infinity.
\end{theorem}

\bigskip
The conformal affine parameter $\bar{\tau}$ defines, in a natural way, 
a foliation of the manifold $\mathring{\mathcal{M}}$. Let
$\mathcal{S}_{\bar{\tau}}$ denote the surfaces of constant
$\bar{\tau}$. For fixed $\bar{\tau}$ one has that $\mathcal{S}_{\bar{\tau}}$ is
diffeomorphic to $\Sphere^3$.  Let $\bar{\tau}_0 \in (0,\bar{T}_0)$
and define
\[
\mathcal{S}_0 \equiv \{ - \bar{\tau}_0\}\times \Sphere^3 , \quad \mathcal{Z} \equiv \{p \in
\mathcal{S}_0 \vert \bar{\Theta} = 0\}.
\]
The set $\mathcal{S}_0$ intersects null infinity in a
\emph{hyperboloidal way}.  Furthermore, let
\[
\mathring{\mathcal{H}} \equiv \{ p\in \mathcal{S}_0 \vert \bar{\Theta} >0\}.
\]
Define
\[
\tau\equiv \bar{\tau} + \bar{\tau}_0, \quad \mathring{\Theta}(\tau) \equiv
\bar{\Theta}(\tau - \bar{\tau}_0)
\] 
so that $\tau =0 $ on $\mathcal{S}_0$ and
\[
\mathring{\Theta} (\tau) =  \bar{\Omega} \left( \left( 1-\frac{\bar{\tau}^2_0}{\bar{\omega}^2} \right) + 2\frac{\bar{\tau}_0}{\bar{\omega}^2}\tau - \frac{1}{\bar{\omega}^2}\tau^2 \right).
\]
The initial value of $\mathring{\Theta}$ on $\mathcal{S}_0$ will be denoted
 by $\mathring{\Omega}$. It can be verified that $\mathring{\Omega}$
 is a boundary defining function. In what follows, let
\[
\mathring{\mathbf{u}}(\tau,x)\equiv \mathbf{u}(\tau-\bar{\tau}_0,x).
\]
The following is an obvious corollary of theorem \ref{Theorem:Local}
---for details of the proof see the analogous construction in \cite{LueVal09a}.

\begin{corollary}
\label{Corollary:HyperboloidalData}
The field $\mathring{\mathbf{u}}(\tau,x)$ implies hyperboloidal data on
$\mathring{\mathcal{H}}$ for the conformal Einstein-Maxwell field
equations. 
\end{corollary}

\subsubsection{Structure of the conformal boundary}
\label{section:structure:of:the:conformal:boundary}

We consider now hyperboloidal data which is ``close'' in some suitable
sense to the hyperboloidal data given by corollary
\ref{Corollary:HyperboloidalData}. Using analogous arguments to the
ones used in \cite{LueVal09b} one can prove the following result.

\begin{proposition}
\label{Proposition:ConformalBoundary}
Given a radiative electrovacuum hyperboloidal initial data set
$(\mathcal{H}, h_{\alpha\beta},K_{\alpha\beta},\Omega)$ sufficiently
close to a reference radiative electrovacuum data $(\mathring{\mathcal{H}},
\mathring{h}_{\alpha\beta},\mathring{K}_{\alpha\beta},\mathring{\Omega})$,
there exists a choice of initial data for the congruence of conformal
curves such that the conformal factor $\Theta$ is 
given by
\begin{equation}
\label{Theta:Hyperboloidal}
\Theta = \Theta_* + \dot{\Theta}_* \tau + \ddot{\Theta}_*\tau^2,
\end{equation}
with 
\[
\Theta_* =\Omega, \quad \dot{\Theta}_*=\langle d, e_0\rangle, \quad
2\Omega\ddot{\Theta}= g^\sharp(d,d)_*.
\]
Furthermore, if the point $i^+\equiv (-\Omega/\dot{\Theta}_*,0,0,0)$
is contained in the development of the initial data, then it is the
unique point at which the conformal factor $\Theta$  
satisfies the (timelike infinity) conditions
\[
\Theta(i^+)=0, \quad \mbox{\emph d}\Theta(i^+)=0, \quad \mbox{Hess\;} \Theta (i^+)
\mbox{ non-degenerate}.
\] 
\end{proposition} 

As it is customary, let $\mathscr{I}$ (null infinity) denote the set of points for
which $\Theta=0$ where the conformal factor is given by
\eqref{Theta:Hyperboloidal}.

\subsubsection{A stability result for purely radiative spacetimes}
\label{section:stability:result}
The information about the conformal boundary of a hypothetical
radiative Einstein-Maxwell spacetime arising from hyperboloidal data
which is contained in Proposition \ref{Proposition:ConformalBoundary}
allows to readily obtain a stability result for a spacetime belonging
to the class arising from Theorem \ref{Theorem:Local}.  The proof of
the following result is similar to that in \cite{LueVal09b} ---see also\cite{Fri88,LueVal09a}.

\begin{theorem}
\label{Theorem:StabilityHyperboloidal}
Let $\mathbf{u}_0=\mathring{\mathbf{u}}_0 + \breve{\mathbf{u}}_0$ be
hyperboloidal initial data for the Einstein-Maxwell conformal field
equations. Given $\tau_+\equiv -\Omega/\dot{\Theta}_*$ and if $\mathring{u}_0$ is
sufficiently small, there exists on $[0,\bar{\tau}_0]\times
\mathcal{S}$ a unique solution
$\mathbf{u}=\mathring{\mathbf{u}}+\breve{\mathbf{u}}$ to the (reduced) conformal
propagation equations \eqref{upsilon:propagation}, \eqref{phi:propagation},
\eqref{varphi:propagation}, \eqref{psi:propagation}
such that the associated congruence of conformal curves contains no
conjugate points in $[0,\tau_+]$.  The solution
$\mathbf{u}=\mathring{\mathbf{u}}+\breve{\mathbf{u}}$ on $\mathcal{D}^+(\mathcal{S})$ implies a
smooth solution $(\mathcal{M}, \tilde{g}) $ to the electrovacuum Einstein field equations
with vanishing cosmological constant, where $\tilde{g}_{\mu\nu} =
\Theta^{-2}g_{\mu\nu}$ with $\Theta$ given by
\eqref{Theta:Hyperboloidal}. The spacetime $(\mathcal{M}, \tilde{g}) $ 
has a conformal boundary given by the set of points for which $\Theta=0 $. The conformal boundary consists
of the set $\mathscr{I}$, which represents future null infinity, and
the point $i^+\equiv(\tau_+,0,0,0) $, which represents timelike
infinity.
\end{theorem}

\medskip
\noindent
\textbf{Remark.} 
The purely radiative
spacetimes used as reference solutions in our analysis are not
perturbations of the Minkowski spacetime. A way of seeing this is to
consider the Newman-Penrose constants of the spacetime. The
Newman-Penrose constants are a set of absolutely conserved quantities
defined as integrals of certain components of the Weyl tensor and the
Maxwell fields over
cuts of null infinity ---see \cite{NewPen65,NewPen68} and
\cite{ExtNewPen69} for the Einstein-Maxwell case. In
\cite{FriSch87} it has been shown that the value of the Newman-Penrose
constants for a vacuum radiative spacetime coincides with the value of the
rescaled Weyl spinor at $i^+$ ---this result can be extended to the
electrovacuum case using the methods of this article. For the radiative spacetimes arising
from the construction of \cite{Sim89} it can be seen that the value of
the Weyl spinor at $i^+$ is essentially the mass quadrupole of the
\emph{seed} static spacetime. It follows, that the Newman-Penrose
constants of the radiative spacetime can take arbitrary values. On
the other hand, for the Minkowski spacetime, the Newman-Penrose
constants are exactly zero, and those of perturbations thereof will be
small. Thus, in this precise sense, our radiative spacetimes are,
generically, not perturbations of the Minkowski spacetime, unless all
the Newman-Penrose constants vanish.

\section{Acknowledgements}
CL was supported by  a research project
grant (F/07 476/AI) of the Leverhulme Trust.  JAVK was funded by an EPSRC Advanced Research Fellowship.


\end{document}